\documentclass[aps,prb,twocolumn,nolongbibliography,superscriptaddress]{revtex4-2}
\let\revappendix\appendix
\usepackage[T1]{fontenc}
\usepackage[utf8]{inputenc}
\usepackage{graphics}
\usepackage{epsfig}
\usepackage{dcolumn}
\usepackage{bm}
\usepackage{amsmath}
\usepackage{dsfont} 
 \usepackage{mathtools}
 \usepackage{amsfonts}
\usepackage{latexsym}
\usepackage{amssymb}
\usepackage{color}
\usepackage{times}
\usepackage[colorlinks=true,allcolors=blue]{hyperref}
\usepackage[normalem]{ulem}
 \usepackage{bbm}
\usepackage{bbold}
\usepackage{tabularx}

 \newcommand{\tb}[1]{}

\newcommand{\bs}{\boldsymbol}

\newcommand{\stkout}[1]{\ifmmode\text{\sout{\ensuremath{#1}}}\else\sout{#1}\fi}

\newcommand{\be}{\begin{equation}}
\newcommand{\ee}{\end{equation}}
\newcommand{\bea}{\begin{eqnarray}}
\newcommand{\eea}{\end{eqnarray}}
\newcommand{\ba}{\begin{align}}
\newcommand{\ea}{\end{align}}

\newcommand{\nn}{\nonumber}

\begin{document}
 \newcommand{\operator}[1]{\ensuremath{ \hat{\mathbf{#1}}}}

\title{Longitudinal coupling between electrically driven spin-qubits and a resonator}

  \author{Sarath Prem}
  \email{sarathprem@magtop.ifpan.edu.pl}
  \author{Pei-Xin Shen}
  \author{Marcin M. Wysoki\'nski}
  \author{Mircea Trif}
  \email{mtrif@MagTop.ifpan.edu.pl}
    \affiliation{International Research Centre MagTop, Institute of
   Physics, Polish Academy of Sciences,\\ Aleja Lotnik\'ow 32/46,
   PL-02668 Warsaw, Poland}
 
\begin{abstract}
At the core of the success of semiconducting spin qubits is the ability to manipulate them electrically, enabled by the spin-orbit interactions. However, most implementations require external magnetic fields to define the spin qubit, which in turn activate various charge-noise mechanisms. Here we study spin qubits confined in quantum dots at zero magnetic fields that are driven periodically by electrical fields and are coupled to a microwave resonator. Using Floquet theory, we identify a well-defined Floquet spin-qubit originating from the lowest degenerate spin states in the absence of driving. We find both transverse and longitudinal couplings between the Floquet spin qubit and the resonator, which can be selectively activated by modifying the driving frequency. We show how these couplings can facilitate fast qubit readout and the implementation of a two-qubit CPHASE gate. Finally, we use adiabatic perturbation theory to demonstrate that the spin-photon couplings originate from the non-Abelian geometry of states endowed by the spin-orbit interactions, rendering these findings general and applicable to a wide range of solid-state spin qubits.
\end{abstract} 
 
\date{\today}
\maketitle
\section{Introduction}

Solid-state systems-based quantum bits (qubits) have seen tremendous progress in the last decades \cite{Laucht_Nanotechnology.2021Roadmap,Gyongyosi_ComputerScienceReview.2019Survey}. 
Among the solid-state platforms, electron and hole spins localized in semiconductor quantum dots (QDs) have been at the forefront of research because of their long coherence times, scalability, and natural compatibility with industrial grade semiconductor manufacturing technologies \cite{Loss_Phys.Rev.A.1998Quantum,Kloeffel_Annu.Rev.Condens.MatterPhys..2013Prospects,Zajac_Phys.Rev.Appl..2016Scalable,Chatterjee_NatRevPhys.2021Semiconductor,Burkard_Rev.Mod.Phys..2023Semiconductor,Warburton_NatureMater.2013Single,Fang_Mater.Quantum.Technol..2023Recent,Piot_Nat.Nanotechnol..2022Single}. 
Fast and high fidelity readout \cite{Harvey-Collard_Phys.Rev.X.2018HighFidelity,Zheng_Nat.Nanotechnol..2019Rapid}, 
single-qubit control  \cite{Yoneda_NatureNanotech.2018Quantumdot,Yang_NatElectron.2019Silicon,Hendrickx_NatCommun.2020Singlehole}, 
and two-qubit logic \cite{Xue_Nature.2022Quantum,Noiri_Nature.2022Fast,Madzik_Nature.2022Precision}
have been demonstrated in different experiments. 

One of the main ingredients that have accelerated the advances in spin-qubit implementations is the strong spin-orbit interaction (SOI) that electrons and holes experience in semiconductors, as it facilitates their fast manipulation by external electrical fields via the electric dipole spin resonance (EDSR)  \cite{Golovach_Phys.Rev.B.2006Electricdipoleinduced,Nowack_Science.2007Coherent,Nadj-Perge_Nature.2010Spin,Pribiag_NatureNanotech.2013Electrical,Corna_npjQuantumInf.2018Electrically,Gao_Adv.Mater..2020SiteControlled,Jirovec_Nat.Mater..2021Singlettriplet}. 
Furthermore, it enables interface spins with photons in microwave resonators (or cavities), which allows fast qubit state detection and generation of long-range spin entanglement  \cite{Blais_Rev.Mod.Phys..2021Circuit,Bonsen_Phys.Rev.Lett..2023Probing,Wallraff_Nature.2004Strong,Yu_Nat.Nanotechnol..2023Strong}. 
The downside of strong SOI, however, is that it exposes the spin to various charge-noise sources, leading to a reduced spin-qubit coherence, in particular in the presence of an external magnetic field that splits the spin states. While that can be mitigated by operating the qubit at sweet spots where the decoherence is minimized 
\cite{Watzinger_NatCommun.2018Germanium,Corna_npjQuantumInf.2018Electrically,Wang_npjQuantumInf.2021Optimal,Bosco_PRXQuantum.2021Hole,Malkoc_Phys.Rev.Lett..2022ChargeNoiseInduced,Michal_Phys.Rev.B.2023Tunable}, 
it requires fine tuning of the electrostatic confinements, which restricts the parameter phase space where they can be operated. 

Driving the spin qubit can enhance its coherence. For example, it was recently shown that by shuttling electrons in an array of $3\times 3$ laterally defined QDs in GaAs heterostructure \cite{Mortemousque_PRXQuantum.2021Enhanced}, 
the coherence time was enhanced by a factor of 10 compared to stationary QDs. This was because the effect of the hyperfine interactions, which dominate the dephasing in the static GaAs QDs, averages to zero during the dynamics, in analogy to the effect of motional narrowing phenomenon observed in NMR  \cite{Slichter_.1978Principles,Huang_Phys.Rev.B.2013Spin}.  
Furthermore, recent works have shown that Floquet qubits encoded in the quasienergy structure occurring in periodically driven superconducting systems can exhibit a larger number of sweet spots for operation than their static counterparts, which facilitates their operation \cite{Mundada_Phys.Rev.Appl..2020FloquetEngineered,Huang_Phys.Rev.Appl..2021Engineering,Gandon_Phys.Rev.Appl..2022Engineering,Nguyen_Nat.Phys..2024Programmablea}.

Dynamics not only extends the spin-qubit coherence but also enables its manipulation in the absence of any externally applied magnetic fields (which are used routinely to define the spin-qubit). Indeed, as demonstrated in Refs.~\cite{San-Jose_Phys.Rev.Lett..2006Geometrical,San-Jose_Phys.Rev.B.2008Geometric,Trif_Phys.Rev.Lett..2009Relaxation}, 
the spin of an electron or hole confined in a QD acquires a non-Abelian geometric phase when it is displaced on closed trajectories by electrical fields in the presence of SOI. It was shown theoretically that, by appropriately choosing the QD trajectory, it is possible to implement a universal set of single-qubit gates \cite{Golovach_Phys.Rev.A.2010Holonomic} 
and even generate entanglement \cite{Wysokinski_Phys.Rev.B.2021Berry}.
Overall, the geometric spin manipulation is potentially more robust than the resonant EDSR, since it is not affected by gate timing errors and charge noise \cite{DeChiara_Phys.Rev.Lett..2003Berry,Carollo_Phys.Rev.Lett..2004Spin,Lupo_Phys.Rev.A.2007Robustness}.

In this paper, we consider the coupling between electrically driven SO coupled QDs hosting electron or hole spins in zero applied magnetic field and a microwave resonator. Using the Floquet theory appropriate for periodic driving, we identify a well-defined Floquet spin qubit encoded in two of the quasienergy levels. We unravel a novel longitudinal coupling between the spin qubit driven at frequency $\Omega$ and the photons
\begin{align}
    H_{s-p}\propto \Omega\,\tau_z^F(a^\dagger+a)\,,
    \label{eq1}
\end{align}
which occurs in the absence of any applied magnetic field, and is activated when $\Omega$ matches the frequency of the resonator. When two such qubits are immersed in the same resonator, we demonstrate that a CPHASE gate can be implemented on fast time scales, in particular for hole-spin qubits. To test its robustness, we use a Floquet-Born-Markov density matrix approach to establish the influence of charge noises, and estimate that the Floquet spin qubits are long-lived. Finally, we show that the coupling to the photons originates from the electron or hole spin qubit geometry of states endowed by the SOI, rendering our proposal general and applicable to a wide variety of spin-qubit implementations.

The paper is organized as follows: in Sec.~\ref{Sec:ModelResults} we introduce the model and review the static spin qubits in QDs coupled to a microwave resonator. Then, in Sec.~\ref{Sec:FloquetPhotonInteractions} we use Floquet theory for periodically driven spins by electrical fields, identify the Floquet qubit, and determine its coupling to the resonator. In Sec.~\ref{Sec:FloquetReadout} we show how to readout its state and spectrum with both longitudinal and transverse interactions, while in Sec.~\ref{Sec:Decoherence} we discuss the coherence of the Floquet spin qubits using the Floquet-Born-Markov theory. In Sec.~\ref{Sec:TwoQubitGates} we show how to construct a CPHASE two-qubit gate and estimate its time scales. In Sec.~\ref{Sec:Geometrical} we offer a geometrical perspective of the combined dynamics and infer the generality of our findings. Finally, in Sec.~\ref{Sec:Outlook} we end up with conclusions and an outlook.

\section{Model and static results}
\label{Sec:ModelResults}

\begin{figure}[t]
\includegraphics[width=0.99\linewidth]{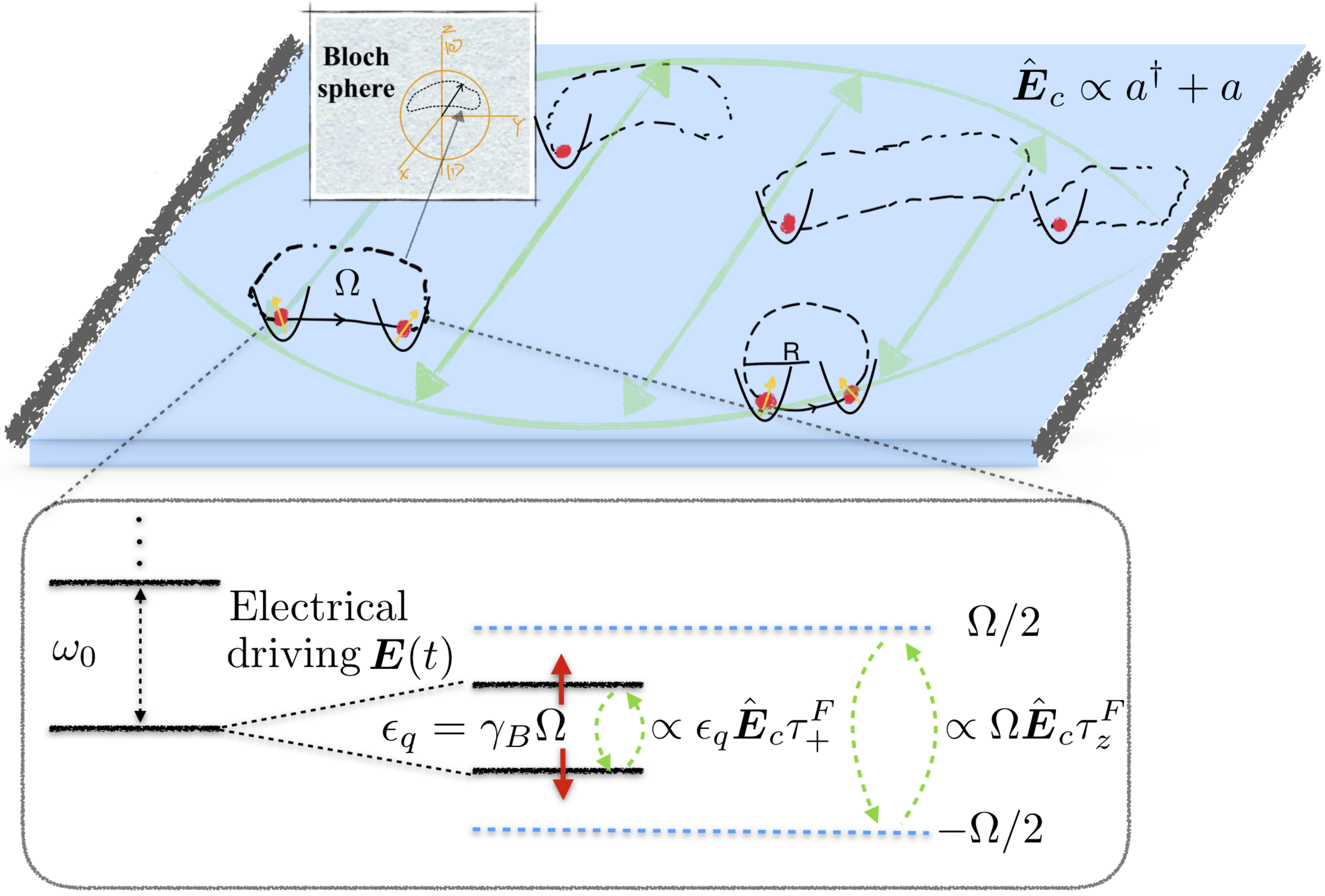}
\caption{Schematics of the setup. Several single electron or hole spins confined in QDs with level spacing $\omega_0$ and coupled to the electric field (green arrows) of a microwave resonator. Applying time-dependent electrical fields ${\bs E}(t)$ oscillating at a frequency $\Omega\ll\omega_0$,  displace the QDs  on closed trajectories in the semiconductor, inducing in turn rotations of the lowest degenerate spinors on the Bloch sphere. The Berry phase accumulated during the rotation, $\gamma_B$,  defines an effective Floquet spin-qubit Hamiltonian with a splitting $\epsilon_q\sim\gamma_B\Omega$. The QD dynamics triggers both transverse ($\propto\epsilon_q\hat{\bs E}_c\tau_+^F$) and longitudinal ($\propto\Omega\hat{\bs E}_c\tau_z^F$) spin-photon interactions, which are activated for $\omega_c\sim\epsilon_q$ and $\omega_c\sim\Omega$, respectively, allowing to detect the Floquet spin qubit state and entangle remote spins interacting with a common resonator.}
\label{fig:1} 
\end{figure}

For concreteness, we apply our dynamical scheme to gate-defined QDs, hosting either electrons or holes. A schematic of the system is shown in Fig.~\ref{fig:1}. It consists of one (or several) electron or hole spins, confined in mobile QDs defined in a two-dimensional semiconductor and immersed in a common microwave resonator with fundamental frequency $\omega_c$. Moreover, the QDs are well separated from each other, so that there is no tunneling between them. The Hamiltonian describing one of the QDs ($\hbar=1$) \cite{Burkard_Rev.Mod.Phys..2023Semiconductor} is 
\begin{align}
     H_{\rm tot}(t)&=H_{0}(t)+H_{SO}+H_{e-p}+\omega_c a^\dagger a\,,\nonumber\\
     H_0(t)&=\frac{{\bs p}^2}{2 m}+U({\bs r})
     +e{\bs r}\cdot{\bs E}(t)\,,\\
     H_{SO}&=\alpha(p_{x}\sigma_{y}\!-\!p_{y}\sigma_{x})\!+\!\beta(\!-p_{x}\sigma_{x}\!+\!p_{y}\sigma_{y})\,,\nonumber\\
     H_{e-p}&=e{\bs r}\cdot{\bs E}_c(a^\dagger+a)\nonumber\,,
 \end{align}  
where ${\bs p}=-i\partial/\partial{{\bs r}}$ is the electron/hole momentum in the QD, 
${\bs E}(t)$ is the classical electrical field acting on the QD,  while ${\bs E}_c$ is the amplitude of the resonator electric field with $a$ ($a^\dagger$) being the annihilation (creation) operator for the resonator photonic mode. Here, $U({\bs r})$ is the confinement potential, while $H_{SO}$ represents the spin-orbit coupling Hamiltonian that accounts for the Rashba ($\propto\alpha$) and Dresselhaus ($\propto\beta$) \cite{Winkler_.2003Spinorbit}. 
Also, $\sigma_{\beta}$ with $\beta=x,y,z$, are the Pauli matrices that act on the electronic spin confined in the QD. Without loss of generality, we assume in the following the parabolic confinement potential for each of the dots, i.e. $U({\bs r})=m\omega_0^2r^2/2$, where $\omega_0$ is the confining frequency. Consequently, in the presence of electric fields $U({\bs r})\rightarrow U[{\bs r}-{\bs R}(t)]$, with ${\bs R}(t)=-e{\bs E}(t)/(m\omega_0^2)$, i.e. a shifted confining potential (up to a constant term of no physical relevance). As demonstrated later in Sec.~\ref{Sec:Geometrical}, more general confining potentials quantitatively alter the resulting Hamiltonians but retain the same qualitative features. Furthermore, other forms of SOI, such as cubic-in-momentum \cite{Winkler_.2003Spinorbit}
can be easily included within the same formalism. \\

This section provides a brief overview of the static spin-orbit qubits that are defined in the presence of a magnetic field. The primary aim of this summary is to establish the baseline results for comparison with the dynamic cases. Throughout the rest of this section we set ${\bs E}(t)=0$, that is, the QDs are not electrically driven.  In the absence of an external magnetic field the QD energy levels are at least doubly degenerate, a consequence of the Kramers theorem. The logical qubit subspace consists of the two lowest eigenstates $\{|\psi_0\rangle,|\psi_1\rangle\}$, w which are split in the presence of an magnetic field ${\bs B}$, by a qubit splitting $\epsilon_q=\epsilon_1-\epsilon_0$ originating from both the Zeeman coupling (via the gyromagnetic factor $g$) and the orbital effects (via the diamagnetic coupling), respectively \cite{Trif_Phys.Rev.B.2008Spin}
. The effective qubit-photon interaction is found by projecting the electron-photon Hamiltonian onto the qubit subspace \cite{Trif_Phys.Rev.B.2008Spin}
which gives (using the identities in Eq.~\eqref{relations} in Appendix \ref{staticcase}): 
\begin{align}
    H_{s-p}&=g_+\tau_+(a^\dagger+a)+{\rm h. c. }\,,\\
    g_+&=i \epsilon_q\frac{{\bs R}_c}{\lambda_{SO}}\cdot{\bs m}_{+}\,,
\end{align}
where ${\bs m}_{+}=\langle\psi_0|{\bs m}|\psi_1\rangle$ with ${\bs m}=\lambda_{SO}{\bs\lambda}_{SO}^{-1}{\bs\sigma}$,
${\bs\lambda}_{SO}^{-1} = \left[\begin{smallmatrix}
    -\beta &\alpha \\
    -\alpha & \beta 
\end{smallmatrix}\right]$
and we defined ${\bs R}_c=e{\bs E}_c/m\omega_0^2$ as the distance over which the electrical field of the resonator shifts the center of the QD. Also, $\tau_\gamma$ are the $\gamma=x,y,z$ Pauli matrices acting in the qubit subspace, with $H_s=\epsilon_q\tau_z/2$ defining the qubit Hamiltonian, while $\lambda_{SO}^{-1}=||{\bs \lambda}_{SO}^{-1}||/\sqrt{2}=m\sqrt{\alpha^2 +\beta^2 } $ represents the Frobenius norm of the SOI matrix \cite{Golovach_Phys.Rev.B.2006Electricdipoleinduced}. 
We can now examine this Hamiltonian in two regimes. Firstly, when the qubit splitting is close to resonance with the photons, $\omega_c\sim \epsilon_q$, we obtain the Jaynes-Cummings Hamiltonian by neglecting the counter-rotating terms. Secondly, in the dispersive regime $|g_+|\ll|\omega_c-\epsilon_q|$, we can determine the qubit state from the resonator frequency shift   \cite{Trif_Phys.Rev.B.2008Spin,Burkard_Rev.Mod.Phys..2023Semiconductor}.  
Moreover, when two of such qubits are coupled to the same resonator, it induces entangling interactions between the two \cite{Trif_Phys.Rev.B.2008Spin,Blais_Rev.Mod.Phys..2021Circuit}
that can be used for the construction of two-qubit gates.  

The spin-photon interaction induced by the SOI exhibits two main features. First, it is exact in the SOI, the only condition being weak electron-photon coupling and $\omega_0\gg \epsilon_q$. Moreover, it vanishes in the limit $B\rightarrow0$ and therefore $\epsilon_q\propto B$, again a consequence of the Kramers theorem. Second, the coupling is purely transverse ($\propto\tau_{x,y}$) in leading order in the electron-photon coupling, which limits its usefulness. Indeed, a transverse coupling hybridizes the resonator and the spin degrees of freedom, which can result in strong qubit decay (via the Purcell effect) and, moreover, the readout is not truly non-demolition \cite{Boissonneault_Phys.Rev.A.2009Dispersive}. 
While such effects can be mitigated by detuning the qubit strongly from the resonator compared to $|g_+|$, it also results in a reduction of the effective coupling and it is restricted to a small number of photons in the resonator \cite{Blais_Rev.Mod.Phys..2021Circuit}.  
A longitudinal coupling instead, of the form in Eq.~\eqref{eq1}, does not suffer from these limitations and can facilitate the readout of the qubit, as demonstrated in Ref.~\cite{Didier_Phys.Rev.Lett..2015Fast}
. Furthermore, it allows for fast and high-fidelity two-qubit gates  \cite{Royer_Quantum.2017Fast,Harvey_Phys.Rev.B.2018Coupling,Bosco_Phys.Rev.Lett..2022Fully}. 
In the following, we turn the discussion to driven QDs and demonstrate the emergence of such longitudinal terms and how they can be selectively activated.

\section{Floquet spin qubit|photon interactions}
\label{Sec:FloquetPhotonInteractions}

EDSR is a well-established technique used to manipulate solid-state qubits, specifically electron or hole spins, that are localized in quantum dots (QDs). Such control necessitates external magnetic fields that induce qubit splitting. Here, we scrutinize whether driving the QDs electrically in the SOI field allows us to define, manipulate, and detect them in the absence of any such magnetic fields. Therefore, in this section we set the external magnetic field $B=0$,
and switch on the time-dependent electrical fields ${\bs E}(t+T)={\bs E}(t)$ acting on a single QD, while in a later section we discuss the implications for several such QDs. The natural theoretical framework for describing the dynamics in this case is the Floquet theory \cite{Rudner_NatRevPhys.2020Band}. 
When the QD is not coupled to the resonator, the time-dependent Schrodinger equation that governs the QD's evolution can be expressed as follows:
\begin{align}
    i\partial_t|\Psi_j(t)\rangle=H_{el}(t)|\Psi_j(t)\rangle\,,
\end{align}
where $H_{el}(t)=H_{0}(t)+H_{SO}$, and the state $|\Psi_j(t)\rangle$, with $j=0,1,2,\dots$, can be expressed as  $|\Psi_j(t)\rangle=e^{-i\epsilon_jt}|\psi_j(t)\rangle$, where $\epsilon_j$ is the Floquet quasienergy [defined ${\rm mod}(2\pi/T))$], and $|\psi_j(t+T)\rangle=|\psi_j(t)\rangle$ is the periodic component of the Floquet state \cite{Giovannini_J.Phys.Mater..2019Floquet}.
They therefore form a complete basis, defining the unique set of stationary solutions to the Schrodinger equation. The corresponding electronic evolution operator reads:
\begin{align}
U(t,0)=\sum_{j}|\psi_j(t)\rangle\langle\psi_j(0)|e^{-i\epsilon_jt}\,,
    \label{evolution}
\end{align}
and,  hence, the matrix element of an operator $\mathcal{A}$ of the system (in the interaction picture) becomes:
\begin{align}
\mathcal{A}(t)=\sum_{j,j'}\mathcal{A}_{jj'}(t)|\psi_j(0)\rangle\langle\psi_{j'}(0)|e^{i(\epsilon_j-\epsilon_{j'})t}\,,
\end{align}
with $|\psi_j(0)\rangle$ being the Floquet states at time $t=0$ when the measurement starts and $\mathcal{A}_{jj'}(t)=\langle\psi_j(t)|\mathcal{A}|\psi_{j'}(t)\rangle$. Furthermore, since $\mathcal{A}_{jj'}(t+T)=\mathcal{A}_{jj'}(t)$, we can write $\mathcal{A}_{jj'}(t)=\sum_k\mathcal{A}_{jj'}(k)e^{ik\Omega t}$, with $\mathcal{A}_{jj'}(k)$ the corresponding Fourier component and $\Omega=2\pi/T$ the precession frequency. Of particular interest is the position operator, which, as demonstrated in the Appendix \ref{dynamiccase}, can be written as: 
\begin{align}
    {\bs r}_{jj'}(t)=\frac{i}{m\lambda_{SO}}\sum_{k}\frac{\epsilon_{jj'}+k\Omega}{\omega_0^2-(\epsilon_{jj'}+k\Omega)^2}e^{-ik\Omega t}{\bs m}_{jj'}(k)\,,
\label{Floquet_mat_elem}    
\end{align}
where $\epsilon_{jj'}=\epsilon_j-\epsilon_{j'}$, and ${\bs m}_{jj'}(k)$ is the $k$-th Fourier component of  ${\bs m}_{jj'}(t)=\langle\psi_j(t)|{\bs m}|\psi_{j'}(t)\rangle$. The expression above readily allows one to express the electron-photon interaction Hamiltonian in the Floquet basis states and, moreover,  to select the two Floquet levels that describe well a qubit. In this work, we focus on driving frequencies $\Omega\ll\omega_0$, so that the dynamics remains confined within the lowest instantaneous doublet [a more rigorous condition that quantifies the adiabatic regime is given by $|\dot{\bs R}|\ll\lambda_0\omega_0$ \cite{San-Jose_Phys.Rev.Lett..2006Geometrical,San-Jose_Phys.Rev.B.2008Geometric}, 
with $\lambda_0=\sqrt{1/m\omega_0}$ being the lateral size of the QD]. Therefore, we can pick from the Floquet ladder the ones that originate from this doublet, a choice that we confirm in a later section by using an adiabatic perturbation theory. 

By projecting the expression in Eq.~\eqref{Floquet_mat_elem} onto the subspace spanned by the aforementioned two Floquet states, $\{|\psi_0(0)\rangle,|\psi_1(0)\rangle\}$ (see Appendix \ref{numerics}), we find:
\begin{align}
    H_{s-p}(t)&=[g_z(t)\tau_z^F+(g_+(t)\tau_+^F+{\rm h. c. })](a^\dagger+a)+\omega_ca^\dagger a\nonumber\,,\\
    g_z(t)&=\frac{{\bs R}_{c}}{2\lambda_{SO}}\cdot\frac{d}{dt}{\bs m}_{z}(t)\label{spin_ph_Fl}\,,\\
    g_+(t)&=ie^{i\epsilon_q t}\frac{{\bs R}_c}{\lambda_{SO}}\cdot\left(\epsilon_q-i\frac{d}{dt}\right){\bs m}_{+}(t)\nonumber\,,
\end{align}
where ${\bs\tau}^F=(\tau_x^F,\tau_y^F,\tau_z^F)$ are Pauli matrices acting in the Floquet subspace spanned by the above states,  $\tau_{\pm}^F=\tau_x^F\pm i\tau_y^F$, and $\epsilon_q=\epsilon_1-\epsilon_0$ is the Floquet qubit splitting. Also, ${\bs m}_z(t)=\langle\psi_1(t)|{\bs m}|\psi_1(t)\rangle-\langle\psi_0(t)|{\bs m}|\psi_0(t)\rangle$ and ${\bs m}_{\pm}(t)=\langle\psi_{1,0}(t)|{\bs m}|\psi_{0,1}(t)\rangle$. Eq.~\eqref{spin_ph_Fl} constitutes one of our main results: it describes the spin-photon coupling for SO-coupled spin qubits in semiconducting QDs that are periodically driven by electrical fields. Inspecting the above expressions, we see that ($i$) the second term is the analogue of the static spin-photon coupling with the Floquet energy $\epsilon_q$ playing the role of the Zeeman splitting, and $(ii)$ the first term, $\propto\Omega$, represents a longitudinal coupling to the resonator and has no analogue for static spin qubits in magnetic fields.  

Let us first analyze in more detail the origins of the qubit splitting. In the case of a Floquet system, the quasienergy associated with a state $j$ can be expressed as $\epsilon_j=\bar{\epsilon}_j-\phi_{G,j}/T$, with \cite{Aharonov_Phys.Rev.Lett..1987Phase,Reynoso_NewJ.Phys..2017Spin}
\begin{align}
    \bar{\epsilon}_j=&(1/T)\int_0^T dt\langle\psi_j(t)|H_{el}(t)|\psi_j(t)\rangle\nonumber\,,\\
    \phi_{G,j}&=i\int_0^T dt\langle\psi_j(t)|\partial_t|\psi_j(t)\rangle\,,
\end{align}
being the average energy and the geometrical phase in the Floquet state $|\psi_j(t)\rangle$, respectively. In the adiabatic limit, the geometric phase becomes the Berry phase, or $\gamma_{B,j}=\phi_{G,j}(T\rightarrow\infty)$. This is a phase factor acquired by the system, which depends solely on the geometry of the path traced out in the parameter space (here, the position of the QD in the 2D semiconductor), rather than the specific details of the evolution. Consequently, when driving is adiabatic, we can define the quasienergies as the instantaneous energies $\bar{\epsilon}_{j}$ modified by the Berry phases, i.e. $\epsilon_{j}=\bar{\epsilon}_{j}-\gamma_{B,j}/T$. Since for the lowest Kramer doublet $\bar{\epsilon}_{0}=\bar{\epsilon}_{1}$, the whole qubit splitting is generated by the Berry phases in leading order in $\Omega$, $\epsilon_q=\gamma_B\Omega+\mathcal{O}(\Omega^2)$, where $\gamma_B\equiv(\gamma_{0,B}-\gamma_{1,B})/2\pi$ and depends solely on the trajectory of the QDs during periodic driving. Consequently, the Fourier components of the Floquet qubit photon coupling in Eq.~\eqref{spin_ph_Fl} oscillate as
\begin{align}
    g_z(t)\sim k\Omega e^{ik\Omega t}\,;\,\,\,\,\,\,\,\,g_+(t)\sim (\gamma_B+k)\Omega e^{i(\gamma_B+ k)\Omega t}
\end{align}
with $k=0,1,\dots$, allowing to selectively activate them by tuning the resonator frequency close to the corresponding frequencies. For example, when $\omega_c\sim\gamma_B\Omega$, we can disregard the term $\propto g_z(t)$, resulting in a purely transverse coupling, as in the static case  \cite{Trif_Phys.Rev.B.2008Spin}.
More interestingly, however, is when $\omega_c\sim\Omega$. Then, by switching to a frame that rotates at a frequency of $\Omega$, which can be achieved through a unitary transformation $\widetilde{U}_{ph}(t) =e^{-i \Omega a^\dagger a t}$, the Hamiltonian becomes:
\begin{align}
    H_{s-p}\approx\Delta a^\dagger a+\frac{\Omega}{2}\frac{{\bs R}_c}{\lambda_{SO}}\cdot{\bs m}_z(k=1)\tau_z^F(a^\dagger+a)\,,
\end{align}
where $\Delta=\omega_c-\Omega$. It is important to note that we made the assumption that $\{\Delta,\,\frac{\Omega}{2}\frac{{\bs R}_c}{\lambda_{SO}}\cdot{\bs m}_z(k=1)\} \ll \{ \omega_c, \Omega \}$, which allows us to ignore all terms that oscillate rapidly on the time scale $1/\Delta$. We mention that in a recent work \cite{Gandon_Phys.Rev.Appl..2022Engineering}
a scheme has been proposed to engineer a longitudinal coupling between Floquet (superconducting) qubits and photons in a microwave resonator. There, for creating such a coupling an additional driving tone was required, on top of the periodic pulse generating the Floquet qubit. Here, instead, such longitudinal coupling is induced by the driving itself, and, moreover, the separation in frequency (which deactivates the transverse coupling) is caused by the accumulated Berry phase during the cyclic motion. Therefore, such a mechanism is intrinsic and does not require additional driving tones, reducing the complexity of implementation.     

\subsection*{Specific drivings}

In order to calculate the Floquet matrix elements in Eq.~\eqref{spin_ph_Fl}, it is helpful to transform to a reference frame that moves with the QD. This is  achieved by the unitary transformation 
$\mathcal{U}[{\bs R}(t)]=e^{-i{\bs p}\cdot{\bs R}(t)}$, so that the isolated QD Hamiltonian Hamiltonian becomes:
\begin{align}
    \widetilde{H}_{el}(t)=H_{el}(0)-{\bs p}\cdot\dot{\bs R}(t)\,,
\label{mf}
\end{align}
and therefore the entire time-dependence stems from the velocity term $\propto\dot{\bs R}(t)$. Note that $[\mathcal{U}({\bs R}),{\bs\sigma}]=0$, and therefore the spin operator ${\bs\sigma}$ is the same in the two frames and allows us to use the wave functions of $\widetilde{H}_{el}(t)$ to calculate the matrix elements. Focusing on the adiabatic regime, we can project Eq.~\eqref{mf} onto the lowest degenerate subspace pertaining to $H_{el}(0)$ to obtain the qubit evolution Hamiltonian \cite{San-Jose_Phys.Rev.B.2008Geometric}
(for more details, see Sec.~\ref{Sec:Geometrical}):
\begin{align}
    \widetilde{H}_s(t)\equiv \mathcal{P}_s\widetilde{H}_{el}(t)\mathcal{P}_s=\frac{1}{\lambda_{SO}}\dot{\bs R}(t)\cdot{\bs m}\,,
    \label{degen_ham}
\end{align}
where $\mathcal{P}_s$ is the projector onto the degenerate subspace. Since this Hamiltonian acts only in spin space, the Floquet states will be simply time-dependent superpositions of the basis states $\{|\uparrow\rangle,|\downarrow\rangle\}$:
    \begin{align}
    |\psi_1(t)\rangle&=\cos\left(\frac{\theta(t)}{2}\right)|\uparrow\rangle+e^{i\phi(t)}\sin\left(\frac{\theta(t)}{2}\right)|\downarrow\rangle\,,\nonumber\\
    |\psi_0(t)\rangle&=e^{-i\phi(t)}\sin\left(\frac{\theta(t)}{2}\right)|\uparrow\rangle-\cos\left(\frac{\theta(t)}{2}\right)|\downarrow\rangle\,,
    \label{Floquet_states}
\end{align}
with the angles $\theta(t)$ and $\phi(t)$ determined by the specific trajectory ${\bs R}(t)$. Corrections to these states caused by interaction with higher orbital levels are proportional to $(\Omega/\omega_0)$ and can be ignored in the adiabatic limit. In the following, we discuss several examples of trajectories that lend themselves to exact analytical solutions: ($i$) linearly, ($ii$) circularly, and ($iii$) elliptically polarized driving, respectively. To simplify the discussion, we will consider only the Rashba spin orbit interaction (SOI) in the following, which implies ${\bs m}=(-\sigma_y,\sigma_x,0)$. 

\subsection{Linear trajectory}

In case ($i$), we assume ${\bs R}(t)={\bs e}_yR_{0,y}\sin(\Omega t)$, with $R_{0,y}$ the displacement amplitude, which gives:
\begin{align}
    H_{s}(t)=-\frac{\Omega R_{0,y}}{\lambda_{SO}}\sigma_x\cos(\Omega t)\,.
\end{align}
The corresponding Floquet quasienergies and states are, respectively, $\epsilon_{1,0}=0$ and \cite{Hausinger_Phys.Rev.A.2010Dissipative}
\begin{align}
    |\psi_{0,1}(t)\rangle=\frac{1}{\sqrt{2}}(|\uparrow\rangle\pm|\downarrow\rangle)\sum_{p\in\mathcal{Z}}e^{\pm ip\Omega t}J_p\left(\frac{R_{0,y}}{\lambda_{SO}}\right)\,,
\end{align}
where $J_p(x)$ is the Bessel function of second kind. This corresponds to a degenerate Floquet qubit state, and thus any combination of the mentioned states is also a legitimate Floquet state. The longitudinal coupling, as defined in Eq.~\eqref{degen_ham}, vanishes $g_z(t)\equiv0$ because the wave-functions only possess global time-dependent phases. However,  when $\Omega\sim\omega_c$, we find the following effective spin-photon coupling
\begin{align}
    H_{s-p}&\approx\Delta a^\dagger a+2\Omega\frac{R_{c,x} R_{0,y}}{\lambda_{SO}^2}\tau_x^F(a^\dagger+a)\,,
\label{linear}    
\end{align}
which, up to a $\pi$ rotation around $\tau_y^F$, is identical to the second term in Eq.~\eqref{spin_ph_Fl}. That is, when the driving is linear, the spin-photon coupling is purely longitudinal and can now be utilized for readout, manipulation, and entanglement of remote Floquet spin qubits (see Appendix \ref{linearpath} for the full general expression). In order for this coupling to emerge, the resonator's electric field must possess a component that is perpendicular to the electric field responsible for inducing the linear motion of the QD. Later, we show that this requirement is related to the crucial role of the non-Abelian Berry curvature associated with the QD motion in the SO-coupled material.

\subsection{Circular trajectory}

In scenario ($ii$), the dot's center undergoes circular motion within the SO-coupled semiconductor, with ${\bs R}(t)=R_0[\cos(\Omega t),\sin(\Omega t)]$, where $R_0$ represents the radius of the trajectory. When inserted into Eq.~\eqref{degen_ham}, it describes a spin $1/2$ in a rotating magnetic field and can be rendered static by switching to a rotating frame with a transformation $V(t)=\exp(i\Omega\sigma_zt/2)$. This gives the following
\begin{align}
    \widetilde{H}_s(t)=\frac{\Omega R_0}{\lambda_{SO}}\sigma_x-\frac{\Omega}{2}\sigma_z\,,
\end{align}
and the corresponding spectrum
\begin{align}
    E_{\pm}=\pm\frac{\Omega}{2}\sqrt{1+\left(\frac{2R_0}{\lambda_{SO}}\right)^2}\equiv\pm\frac{\Omega_R}{2}\,.
\end{align}
Returning to the lab frame, we can determine the Floquet states using Eq.~\eqref{Floquet_states}, with the angle $\theta=\arccos{(-\Omega/\Omega_R)}$ and $\phi(t)=\Omega t$, while the Floquet qubit splitting is $\epsilon_{q}=\Omega_R-\Omega$. We find that when $\Omega\sim\omega_c$ the resulting coupling is longitudinal:
\begin{align}
    H_{s-p}&=\Delta a^\dagger a+2\Omega\frac{R_0R_{c}}{\lambda_{SO}^2}\left(a^\dagger+a\right)\tau_z^F\,,
    \label{long_circ}
\end{align}
where $R_{c}=\sqrt{R_{c,x}^2+R_{c,y}^2}$ and we assumed $R_0\ll\lambda_{SO}$. On the other hand, when the resonator is in resonance with the Floquet qubit, $\omega_c\sim\epsilon_q$:
\begin{align}
    H_{s-p}&\approx\Delta a^\dagger a+2\Omega\frac{R_c}{\lambda_{SO}}\left(\frac{R_0}{\lambda_{SO}}\right)^2\left(\tau_+^Fa+a^\dagger\tau^F_-\right)\,,
    \label{trans_circ}
\end{align}
which emulates the well-known Jaynes-Cummings model with all its implications  \cite{Haroche_.2006Exploring}.
It should be noted that $R_0$ can, in principle, be arbitrarily large, and therefore the only small parameter is $R_c/\lambda_{SO}$ (while $\Omega\ll\omega_0$ is assumed throughout). 
\begin{figure}[t]
\includegraphics[width=\linewidth]{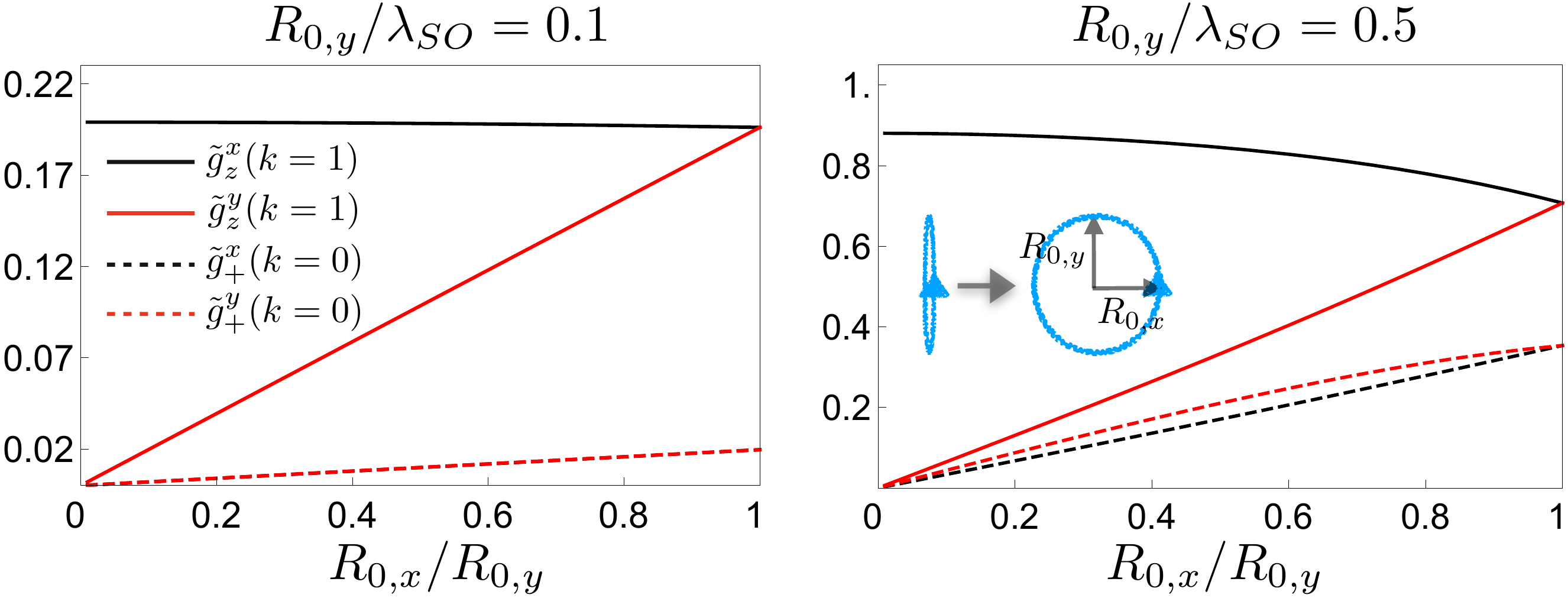}
\caption{Variation of the longitudinal, $\tilde{g}_z^{x,y}(k=1)\equiv m_z^{x,y}(k=1)$,  and transverse $\tilde{g}_{+}^{x,y}(k=0)=\gamma_B m_{+}^{x,y}(k=0)$ spin-photon coupling strengths with the ellipticity $R_{0,x}/R_{0,y}$  for $R_{0,y}/\lambda_{SO}=0.1$ (left) and $R_{0,y}/\lambda_{SO}=0.5$ (right). }
\label{fig:2} 
\end{figure}

\subsection{Elliptical trajectory}

For a general trajectory, ${\bs R}(t+T)={\bs R}(t)$,  the Floquet spectrum and wave functions can only be evaluated numerically. To test cases that interpolate between the two particular drives discussed above, we have evaluated the spin-photon coupling for elliptic trajectories:
\begin{align}
    {\bs R}(t)=(R_{0,x}\cos{\Omega t},R_{0,y}\sin{\Omega t})\,,
\end{align}
where $R_{0,x(y)}$ are the arbitrary amplitudes in the $x$ ($y$) direction. Indeed, for $R_{0,x}=0$ ($R_{0,x}=R_{0,y}$) the motion is linear (circular). Specifically, we have calculated numerically the Fourier transforms of the functions $\tilde{\bs g}_z(t)=\dot{\bs m}_z/\Omega$ and $\tilde{\bs g}_+(t)=(\epsilon_q-d/dt){\bs m}_+/\Omega$ that determine the strength of the spin-photon coupling in Eq.~\eqref{spin_ph_Fl}.  In Fig.~\ref{fig:2} we show the Fourier components $\tilde{g}_z^{x,y}(k=1)$ and $\tilde{g}_+^{x,y}(k=0)$ as functions of $R_{0,x}/R_{0,y}$ for various values of $R_{0,y}/\lambda_{SO}$. Since $\tilde{g}_+^{x,y}(k=0)\propto\gamma_B$, both vanish for a linear drive and increase monotonically toward the circular trajectory where they obey $\tilde{g}_+^{x}(k=0)=\tilde{g}_+^{y}(k=0)$. Importantly, longitudinal couplings dominate over transverse ones in the range $0\leq R_{0,x}/R_{0,y}\leq1$, and therefore can be used for various quantum operations. When dealing with general potentials and drivings in the adiabatic limit, the appropriate approach is the adiabatic perturbation theory, as described in Sec.~\ref{Sec:Geometrical}.

\subsection{Estimates}

The strength of the spin-photon coupling is determined by the $R_{c}/\lambda_{SO}$, and to provide estimates it is instructive to express this ratio as follows:
\begin{align}
\frac{R_{c}}{\lambda_{SO}}\equiv\frac{eE_c\lambda_0}{\omega_0}\frac{\lambda_0}{\lambda_{SO}}\,.
\end{align}
Moreover, for simplicity, we fix $R_0/\lambda_{SO}=0.25$ (which can be achieved by simply changing the driving amplitude depending on the system) and assume $\Omega=\omega_c/ 2\pi=5$ GHz, which represents a typical value for microwave resonators. We first focus on the hole spins, which exhibit large SOI, and choose an overall ratio $\lambda_{0}/\lambda_{SO}=0.2$. For Ge hole-spin qubits encoded in squeezed dots
in Ge/SiGe heterostructures \cite{Bosco_Phys.Rev.Lett..2022Fully}
(confining length of $\lambda_0=50$ nm and $\omega_0/ 2\pi= 5.2\times10^{2}$ GHz) results in a ratio $R_c/\lambda_{SO}\sim10^{-3}$ and a coupling strength $g_z\sim 2$ MHz, while for Ge/Si core/shell nanowires ($\lambda_0= 25$ nm and $\omega_0/ 2\pi= 9.8$ GHz) QDs we find $R_c/\lambda_{SO}\sim10^{-2}$ and $g_z\sim 20$ MHz. Also, for Si hole-spin qubits encoded in square
fin field-effect transistors (FETs) with $\lambda_0=20$ nm and $\omega_0/2\pi= 2\times10^{2}$ GHz \cite{Bosco_Phys.Rev.Lett..2022Fully},  
$R_{c}/\lambda_{SO} \approx 1.9 \times 10^{-3}$ and $g_z\sim 4$ MHz, for a confinement frequency $\omega_0\sim2.4\times 10^2$ GHz. We note that the SOI can be strongly altered by electrical fields in the case of holes \cite{Kloeffel_Phys.Rev.B.2011Strong,Kloeffel_Phys.Rev.B.2013Circuit,Wang_npjQuantumInf.2021Optimal,Bosco_Phys.Rev.Lett..2022Fully}. 
For electrons confined in typical GaAs QDs, $\lambda_0=50$ nm $\lambda_0/\lambda_{SO}\sim 0.1$, giving  $R_{c}/\lambda_{SO} \approx 5 \times 10^{-5}$ and  $g_z\sim 0.1$ MHz. Although smaller than in the case of holes, this coupling can be increased considering cavities with higher frequencies.\cite{Burkard_Rev.Mod.Phys..2023Semiconductor}
For $\omega_c\sim 50$ GHz, the coupling reaches MHz values, comparable to the (transverse) static spin-photon couplings \cite{Petersson_Nature.2012Circuit,Mi_Nature.2018Coherent}. 

\section{Floquet spin qubit readout}
\label{Sec:FloquetReadout}

Both the strength of the Floquet qubit-photon coupling and the Floquet spectrum can be extracted from the response of the resonator. The equation of motion for the photon field $a(t)$ for a one-sided resonator coupled to an external line reads  \cite{Clerk_Rev.Mod.Phys..2010Introduction}:
\begin{align}
    \dot{a}=-i\omega_ca+ig_z(t)\tau_z^F-i(g_{-}(t)\tau_+^F+{\rm h. c. })-\frac{\kappa}{2}a-\sqrt{\kappa}a_{in}\,,\label{input}
\end{align}
where $\kappa$ is the escape rate of the resonator, $a_{in}$ is the stream of photons impinging on the resonator, and all operators are in the Heisenberg picture. Furthermore, the output field that exits the resonator, which is eventually detected,  satisfies $a_{out}=a_{in}+\sqrt{\kappa}a$. 

\subsection{Longitudinal coupling readout}

First, we examine the state of the resonator when $\Omega\sim\omega_c$ and the input field is the vacuum. In this case, we can retain only the $g_z(t)$ term in the above equation and, more specifically, its $k=\pm1$ components. In the semiclassical regime, the average $\alpha(t)=\langle a(t)\rangle$ becomes  \cite{Didier_Phys.Rev.Lett..2015Fast}:
\begin{align}
    \alpha(t)=-i\frac{g_z(k=1)\langle\tau_z^F\rangle}{\kappa}(1-e^{-\kappa t/2})\,,
\end{align}
resulting in a finite photon population induced by the time-dependent $g_z(t)$. The effect of the qubit on the resonator can be characterized by the measurement pointer state separation $D(t)=|\alpha_+(t)-\alpha_-(t)|$  \cite{Gandon_Phys.Rev.Appl..2022Engineering},  
and the signal-to-noise ratio (SNR):
\begin{align}
    {\rm SNR}(t)=\frac{\sqrt{8}|g_z(k=1)|}{\kappa}\sqrt{\kappa t}\left[1-\frac{2}{\kappa t}(1-e^{-\kappa t/2})\right]\nonumber\,.
\end{align}
In the stationary limit $D(\infty)=2g_z(k=1)/\kappa$ which corresponds to an intra-cavity average photon number $n_{ph}=(2g_z(k=1)/\kappa)^2$. To provide some estimates, let us consider a linearly driven QD with amplitude $R_{0,y}/\lambda_{SO}=0.25$, so that
\begin{align}
    n_{ph}\approx\left( Q\frac{eE_c\lambda_0}{\omega_0}\frac{\lambda_0}{\lambda_{SO}}\right)^2\,,
\end{align}
with $Q=\omega_c/\kappa$ being the cavity quality factor. Using the same QD systems discussed in the previous section, and assuming state-of-the-art resonators with $Q\sim10^5$, the average photon number for Ge hole-spin qubits encoded in squeezed dots in Ge/SiGe heterostructures is \cite{Bosco_Phys.Rev.Lett..2022Fully}
$n_{ph}\approx 10^{2}$,
while for Ge/Si core/shell nanowires \cite{Bosco_Phys.Rev.Lett..2022Fully} 
$n_{ph} \approx 10^{6}$. For Si hole-spin qubits encoded in square fin (FETs) \cite{Bosco_Phys.Rev.Lett..2022Fully}
$n_{ph} \approx 10^{4}$, while for electron spin qubits encoded in gate-defined GaAs QDs $n_{ph} \approx 10^{2}$. Such photonic populations are easily distinguishable by homodyne detection in current experiments. 

\subsection{Transverse coupling readout}

When $\epsilon_q\sim\omega_c$, we can retain in Eq.~\eqref{input} only the transverse coupling, which facilitates a dispersive readout of the qubit.  Although this type of measurement is well established  \cite{Clerk_Rev.Mod.Phys..2010Introduction},  
there has been relatively little focus on the specific case of periodic driving. As a result, we will provide a brief overview of the methodology here \cite{Kohler_Phys.Rev.Lett..2017Dispersive,Trif_Phys.Rev.Lett..2019Braiding,Rudner_NatRevPhys.2020Band}. 
Let us assume that the resonator is driven coherently at the input port with $\epsilon(t)=\epsilon_d\exp(i\omega t)$, with $\omega$ being the driving frequency and $\epsilon_d$ its amplitude. Then, the input field can be characterized by its mean $\alpha_{in}(t)=\langle a_{in}(t)\rangle=\epsilon(t)/\sqrt{\kappa}$ \cite{Clerk_Rev.Mod.Phys..2010Introduction}.
By treating the term $\propto g_+(t)$ in second order in a Dyson-type series expansion (see Appendix \ref{inputoutputdetails} for more details of the derivation), we find that the average cavity field obeys the following equation:
\begin{align}
\dot{\alpha}(t)\approx-i[\omega_c+\chi_\alpha(\omega_c,t)]\alpha(t)-\frac{\kappa}{2}\alpha(t)-\sqrt{\kappa}\alpha_{in}(t)\nonumber\,,
\label{in_out}
\end{align}
where $\chi_\alpha(\omega_c,t)=\int dt'e^{i\omega_ct}\chi_\alpha(t',t)$ with:
\begin{align}
\chi_{\alpha}(t',t)&=-i\theta(t-t')E^2_{c,\alpha}\langle[r_\alpha^I(t'),r_\alpha^I(t)]\rangle\,,
\end{align}
being the time-dependent susceptibility of the isolated system. Here, $\langle\dots\rangle$ represents the trace over the density matrix of the electrons in the absence of the coupling to photons. 
 In the case of periodic driving, the susceptibility obeys $\chi_{\alpha}(t',t)=\chi_{\alpha}(t'+T,t+T)$, and can be written formally as  \cite{Kohler_Phys.Rev.Lett..2017Dispersive,Trif_Phys.Rev.Lett..2019Braiding} : 
\begin{align}
    \chi_{\alpha}(t-\tau,t)=\sum_{q\in\mathcal{Z}}\frac{1}{2\pi}\int d\omega e^{-iq\Omega t-i\omega\tau}\chi_{\alpha}^q(\omega)\,.
\end{align}
Furthermore, in the regime of a good cavity, $\kappa\ll\omega_c,\Omega$, we can retain only the term $q=0$ in the resulting Fourier transform of Eq.~\eqref{in_out} when determining the cavity response. Let us assume that the qubit density matrix can be written as $\rho_S=\sum_{\tau}\rho_{\tau}|\psi_{\tau}(0)\rangle\langle\psi_{\tau}(0)|$, with $\rho_{\tau}$ being the weight of the Floquet state $\tau=\pm$. This includes situations where the qubit is in the pure state $\tau=\pm$ (with $\rho_{\tau}=1$). As shown in the next section, it also represents the steady state of the qubit in the presence of dissipation. When $\omega\sim\epsilon_q$, we find the following
\begin{align}
\chi_{\alpha}^0(\omega)&\approx\epsilon_q^2\left(\frac{R_{c,\alpha}}{\lambda_{SO}}\right)^2|m_{+}^{\alpha}(k=0)|^2\frac{\rho_{+}-\rho_{-}}{\omega-\epsilon_q+i\delta}\nonumber\,,
\end{align}
where $\delta$ quantifies the linewidth of the Floquet levels due to their coupling with the environment. Therefore, the cavity frequency is shifted by a term dependent on the state of the qubit, while altering its $Q$ factor allows one to extract the decoherence rate of the qubit $\delta$  \cite{Blais_Rev.Mod.Phys..2021Circuit}. 
Note that $\epsilon_q\approx\gamma_B\Omega$, and therefore the response vanishes when $\gamma_B\rightarrow0$, which is consistent with the result for linearly polarized driving. Experimentally, the influence of the qubit on the resonator can be examined from changes in the resonator transmission as a function of the driving frequency $\omega$ of the coherent input state \cite{Trif_Phys.Rev.Lett..2019Braiding}.

There are numerous advantages to utilizing longitudinal spin-photon interactions for qubit readout instead of transverse ones, as illustrated in Ref.~\cite{Didier_Phys.Rev.Lett..2015Fast}
 . Most importantly, the longitudinal coupling enables a faster readout of the qubit's state compared to that allowed by the transverse interactions due to the better separation the pointer states. Additionally, longitudinal couplings do not diminish the qubit coherence due to the Purcell effect, unlike what happens during transverse readout. Considering these advantages, we underscore the importance of longitudinal measurement for our Floquet spin-qubit.

\section{Decoherence due to charge fluctuations }
\label{Sec:Decoherence}

While the Floquet description enables the determination of the system's spectral characteristics, it ultimately corresponds to an out-of-equilibrium situation. To establish the coherence of the Floquet spin qubit, we evaluated the effects of bosonic environments, such as phonons and charge fluctuations, on its dynamics.  We assume that the bath couples with the QD via a dipolar Hamiltonian \cite{San-Jose_Phys.Rev.Lett..2006Geometrical}
:
\begin{align}
    H_{e-b}=\sum_{\alpha=x,y}\tilde{r}_\alpha \mathcal{E}_\alpha\,,
    \label{el-b}
\end{align}
where $\tilde{r}_\alpha=r_\alpha/\lambda_0$, and $\mathcal{E}_{\alpha}$ with $\alpha=x,y$ being the fluctuating electric fields stemming from the environment  \cite{Khaetskii_Phys.Rev.B.2000Spin,Golovach_Phys.Rev.Lett..2004PhononInduced,Wang_npjQuantumInf.2021Optimal}.  
For $\Omega\ll\omega_0$, the dynamics is restricted to the lowest doublet, and we can project Eq.~\eqref{el-b} into the Floquet qubit subspace. In the interaction picture, this results in an effective Hamiltonian $H_{s-b}^I(t)=\sum_{\alpha} A_\alpha(t) \mathcal{E}_\alpha(t)$, with 
\begin{align}
A_\alpha(t)&=\sum_{k\in\mathbb{Z}}
\left[
\tilde{r}^z_{\alpha}(k)\tau_z^F +\left(\tilde{r}_{\alpha}^+(k)\tau_+^Fe^{i\epsilon_qt}
+\rm{h.c. }\right)
\right]e^{ik\Omega t}\nonumber\,,
\end{align}  
where $\tilde{r}_{\alpha}^z(k)=[\tilde{r}_\alpha^{11}(k)-\tilde{r}_\alpha^{00}(k)]/2$ and $\tilde{r}_{\alpha}^+(k)=\tilde{r}_\alpha^{10}(k)=[\tilde{r}_\alpha^{01}(-k)]^*$. The density matrix that describes only the qubit can be written as $\rho_S(t)={\rm Tr}_{B}[\rho_{tot}(t)]$, with $\rho_{tot}(t)$ being the total density matrix of the combined system, and the trace was taken over the environment. Starting from the von Neumann equation obeyed by the total density matrix $\dot{\rho}_{tot}(t)=-i[H_{tot}(t),\rho_{tot}(t)]$, we can eliminate the bath using the standard Born-Markov approach \cite{Breuer_.2002Theory}.
The equation of motion for the reduced density matrix becomes the following:
\begin{align}
  \dot{\rho}_S(t)\!=
  \!-\!\sum_{\alpha \beta}\int_{-\infty}^tdt'J_{\alpha\beta}(t-t')[A_\alpha(t),A_\beta(t')\rho_S(t')]\!+\!{\rm h. c. }
    \label{FQ_MEq}
\end{align}
where 
\begin{equation}
    J_{\alpha\beta}(t-t')={\rm Tr}_B[\mathcal{E}_\alpha(t) \mathcal{E}_\beta(t')\rho_B]\delta_{\alpha\beta}\equiv J_{\alpha}(t-t')\,,
\end{equation}
with the trace taken over the bath degrees of freedom described by the thermal density matrix $\rho_B$. Its Fourier transform, denoted by $J_\alpha(\omega)=(1/2\pi)\int_{-\infty}^\infty dtJ_\alpha(t)e^{-i\omega t}$, is known as the bath spectral function and plays a crucial role in the description of the dynamics of the qubit \cite{Leggett_Rev.Mod.Phys..1987Dynamics}.

Substituting $\rho_S(t')\approx\rho_S(t)$ on the right hand side of Eq.~\eqref{FQ_MEq}, which amounts to performing also the Markov approximation \cite{Breuer_.2002Theory},
allows us to express the dynamics $\rho_S(t)$ in a Lindblad form  (see Appendix \ref{Floquet-Born-Markov} for details):
\begin{align}
\dot{\rho}_S(t)&=\sum_{s=\pm,z}\Gamma_s\mathcal{D}_s[\rho_S(t)]\,,\\    \mathcal{D}_s[\rho_S(t)]&=\tau_s^F\rho_S(t)(\tau_s^F)^\dagger-\frac{1}{2}\{(\tau_s^F)^\dagger\tau_s^F,\rho_S(t)\}\,,
    \label{Lindblad}
\end{align}
where $\mathcal{D}_s[\rho_S(t)]$ is the dissipator associated with absorption ($s=+$), emission ($s=-$), and dephasing ($s=z$) processes, respectively, while 
\begin{align}
    \Gamma_{\pm}&=\left(\frac{\Omega\lambda_0}{\omega_0\lambda_{SO}}\right)^2\sum_{\alpha,k}|m_{\pm,\alpha}(k)|^2(k\pm\gamma_B)^2 J_{\alpha}[\Omega(k\pm\gamma_B)]\nonumber\,\\
    \Gamma_z&=\left(\frac{\Omega\lambda_0}{\omega_0\lambda_{SO}}\right)^2\sum_{\alpha,k}|m_{z,\alpha}(k)|^2k^2 J_\alpha(k\Omega)\,,
    \label{rates}
\end{align}
are the corresponding rates. Above, we neglected a small Lamb shift in the Floquet spectrum $\propto\Omega^2$ caused by the bosonic bath. Although the detailed derivation is presented in the Appendix \ref{Floquet-Born-Markov}, a few comments are in order. To obtain Eq.~\eqref{Lindblad}, we have performed the secular approximation, that is, we neglected the terms $\propto e^{\pm2i\epsilon_q t}$ while we retained only the contributions with $q=-k$ in the terms $\propto e^{\pm(k+q)\Omega t}$ in Eq.~\eqref{FQ_MEq}. This strategy is appropriate if both the minimal quasienergy difference and the drive frequency $\Omega$ are much higher than the inverse of the resulting decoherence rates \cite{Breuer_.2002Theory}.
Since in our case $\Gamma_{\pm,z}\propto\Omega^2$, this condition is satisfied automatically in the weak coupling regime, as opposed to periodic drivings induced by $ac$ magnetic fields \cite{Breuer_.2002Theory,Blattmann_Phys.Rev.A.2015Qubit,Kohler_Phys.Rev.Lett..2017Dispersive}. 
We can identify the relaxation and dephasing times of the Floquet qubit as $1/T_1=\Gamma_++\Gamma_-$ and $1/T_2=1/(2T_1)+\Gamma_z$, respectively. Furthermore, the stationary populations of the states  $\rho_\pm=\Gamma_\pm/(\Gamma_++\Gamma_-)$  do not obey the detailed balance condition $\rho_-/\rho_+\neq\exp{(\Omega/k_B{\rm T})}$, where T is the temperature. Nevertheless, the density matrix is diagonal in the Floquet basis (in the Schrodinger picture).
\begin{align}
\rho_S(t)=\sum_{\tau=\pm}\rho_\tau|\psi_\tau(t)\rangle\langle\psi_\tau(t)|\,,
\end{align}
allowing to initialise the Floquet qubit using the dissipation. These expressions can be contrasted with those describing a static Zeeman-split spin qubit \cite{Khaetskii_Phys.Rev.B.2000Spin,Golovach_Phys.Rev.Lett..2004PhononInduced}: 
\begin{align}
    \Gamma^s_{\pm}&= \left(\frac{\epsilon_q\lambda_0}{\omega_0\lambda_{SO}}\right)^2\sum_{\alpha=x,y}|m_{+,\alpha}|^2J_{\alpha}(\pm \epsilon_q)\,,
\end{align}
while $\Gamma_z=0$.  We can readily identify $\Omega$ as analogous to $\epsilon_q$ in the expression of the rates, and given the analogy, suggests that we can infer the decoherence of the Floquet spin qubit from its static counterpart. 

\begin{figure}[t]
\includegraphics[width=\linewidth]{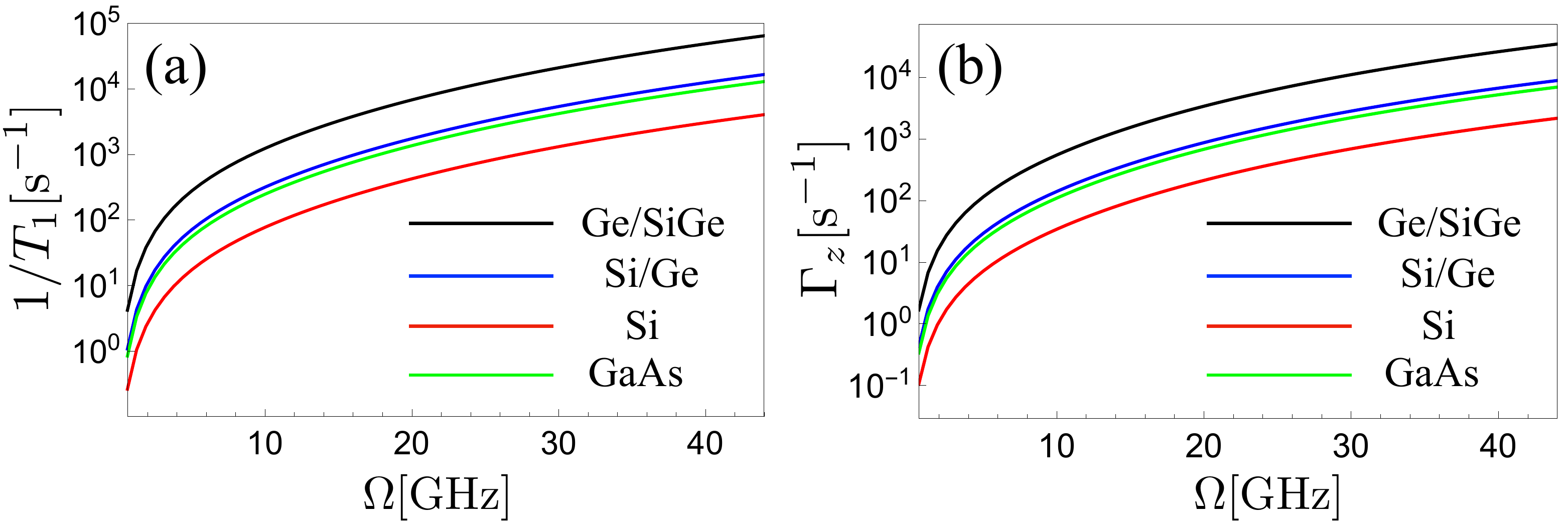}
\caption{The relaxation (left) and the pure dephasing (right) rates as function of the driving frequency $\Omega$ for holes and electron spins confined in various semiconducting QDs. We have used $\lambda_0/\lambda_{SO}=0.2$ for holes (black, red, and blue), while $\lambda_0/\lambda_{SO}=0.1$ for electrons (green). For all plots we assumed $R_0/\lambda_{SO}=0.25$, and T$=40$ mK.}
\label{fig:3} 
\end{figure}

To estimate the decoherence times of the Floquet spin-qubit, we first need to specify the spectral function $J_\alpha(\omega)$. It is convenient to write it as $J_\alpha(\omega)=\rho(\omega)\coth\left(\frac{\omega}{2{\rm T }}\right)$ with $\rho(\omega)$ being the spectral density of the bath. For an ohmic bath at low frequencies $\rho(\omega)=\gamma_\Omega\omega$, with $\gamma_\Omega\sim(e^2/\hbar){\rm Re}[Z]$ being related to the real part of the circuit impedance $Z$. To give estimates, we assume the same QD parameters as in Sec.~\ref{Sec:FloquetPhotonInteractions}, at $\Omega=\omega_c$,  with  a bath temperature T$=40$ mK and $\gamma_\Omega = 10^{-4}$ \cite{San-Jose_Phys.Rev.Lett..2006Geometrical}.
In Fig.~\ref{fig:3} we show the relaxation rate $T_1^{-1}$ (left) and the pure dephasing rate $\Gamma_z$ (right) for Ge hole spin encoded in squeezed dots, Ge / Si core / shell nanowire QDs, hole spin encoded in square
fin Si FET QDs, and for electron spins in gate-defined GaAs, respectively.  We see that these rates stemming from ohmic noises are orders of magnitude lower than the coupling strengths $|g_{z,+}|$ for all of these systems, underlining the robustness of the Floquet spin qubit.

To summarize this section, we identified three features associated with dissipation: ($i$) the coupling between the environment and the qubit is activated only in the presence of driving, which allows us to perform the secular approximation ($ii$) the stationary density matrix is diagonal in the Floquet basis ($iii$) the time-dependent driving induces pure dephasing terms without an analogue in the static Zeeman-split spin qubits.

\section{Two qubit gates with Floquet spin qubits}\
\label{Sec:TwoQubitGates}

When two Floquet spin-qubits are coupled to the same resonator field, the resonator can mediate entangling interactions between the two. Depending on the frequency of the resonator, such an entanglement can be generated by transverse \cite{Blais_Rev.Mod.Phys..2021Circuit}
or longitudinal \cite{Royer_Quantum.2017Fast} 
spin-spin interactions. Here, we only discuss the latter mechanism.   

Modulating simultaneously the longitudinal coupling between a resonator and qubits allows to implement a high-fidelity  controlled-phase gate (CPHASE) \cite{Roos_NewJ.Phys..2008Ion,Royer_Quantum.2017Fast,Leung_Phys.Rev.Lett..2018Robust}.  
We show how to realize such a gate for the Floquet spin-qubits where such modulation occurs naturally as a consequence of the Floquet driving itself. 

We start by assuming that both QDs are driven with the same frequency $\Omega\ll\omega_0$, but can have arbitrary trajectories in the two-dimensional semiconductor (i.e. different Floquet energies $\epsilon_{q,j}$). Assuming $\Omega\sim\omega_c$,  from Eq.~\eqref{spin_ph_Fl} we conclude that the minimum requirements for the longitudinal coupling to dominate over the transverse ones in this limit are $|\Omega-\omega_c|\ll|\epsilon_{q,j}-\omega_c|,|\epsilon_{q,j}+\Omega-\omega_c|$, where $\epsilon_{q,j}$ lies within the first Floquet Brillouin zone $-\Omega/2<\epsilon_{q,j}<\Omega/2$. Interestingly, such a requirement is not present when the QDs are linearly driven, as discussed in the previous section. Then, by retaining only the longitudinal terms, we find the two-qubit Hamiltonian:
\begin{align}
    H_{2q}=\Delta a^\dagger a+(g_{z,1}\tau_{z,1}^F+g_{z,1}\tau_{z,2}^F)(a^\dagger+a)\,,
\end{align}
where $\Delta=\omega_c-\Omega$ (the photons are described in a frame rotating with $\Omega$), and $g_{z,j}\equiv g_{z,j}(1)$  are the $k=1$ components of the longitudinal couplings for the dot  $j=1,2$. Next, following the exposition in~\cite{Harvey_Phys.Rev.B.2018Coupling}
, the evolution operator can be written as:
\begin{align}
U(t)&=e^{i\Delta a^\dagger at}D[\alpha(t)]e^{iJ(t)\tau_{z,1}^F\tau_{z,2}^F}\,,\\
J(t)&=\frac{2g_{z,1}g_{z,2}}{\Delta^2}[\Delta t-\sin(\Delta t)]\,,
\end{align}
while $D[\alpha(t)]=e^{\alpha(t)a^\dagger-\alpha^*(t)a}$ is a displacement operator that depends on both qubits: 
\begin{equation}
    \alpha(t)=\frac{1-e^{i\Delta t}}{\Delta}(g_{z,1}\tau_{z,1}^F+g_{z,2}\tau_{z,2}^F)\,.
\end{equation}
Modulating the couplings for a time $t_g$ that satisfies $\Delta \cdot t_g=2n\pi$, with $n=1,\dots$ and $\frac{2g_{z,1}g_{z,2}}{\Delta}t_g=\pi/4$ decouples the photons from the spins in the dynamics and implements the entangling CPHASE gate $U_{CP}={\rm diag} [1,1,1,-1]$.  This in turn is equivalent to choosing the de-tuning $\Delta=4\sqrt{ng_{z,1}g_{z,2}}$ and the gate time
\begin{equation}
    t_g=\frac{\pi}{2}\sqrt{\frac{n}{g_{z,1}g_{z,2}}}\approx\frac{T}{4}\left(\frac{eE_c\lambda_0}{\omega_0}\frac{\lambda_0}{\lambda_{SO}}\frac{R_0}{\lambda_{SO}}\right)^{-1}\,,
\end{equation}
where in the last line, for simplicity,  we assumed $g_{z,1}=g_{z,2}$. Therefore, the gate time can be decreased by driving the qubits faster, while the duration of the pulse can be controlled by turning on and off the amplitude of the drive $R_0\equiv R_0(t)$. Importantly, the CPHASE gate operation as described above remains the same even if the oscillator is initially in a coherent or thermal state, lifting the requirement of any preparation of the oscillator at the start of the gate \cite{Royer_Quantum.2017Fast,Bosco_Phys.Rev.Lett..2022Fully}.

Targeting specific pairs of qubits is crucial to achieve scalability, and the Floquet spin-qubit  inherently facilitates this capability. In the absence of the driving, the Floquet spin-qubits do not interact with the electric field of the cavity, a consequence of the Kramers theorem. Therefore, only the driven qubits can interact with the cavity and with each other. By driving only the desired pairs, one can selectively couple them while keeping the rest of the QDs idle. For example, this selective coupling enables the implementation of a two-qubit CPHASE gate between the Floquet qubits discussed above.

Let us give some estimates for the gate operation time assuming $\Omega =\omega_c = 2 \pi \times 5 $ GHz, $Q=10^5$ with all other parameters the same as in Sec.~\ref{Sec:FloquetPhotonInteractions}. For Ge hole-spin qubits encoded in squeezed dots in Ge/SiGe heterostructures, we find $t_g\approx 0.7 \mu$ s, for Ge/Si core/shell nanowire QDs $t_g \approx 6.5$ ns, while for Si hole-spin qubits encoded in square fin  (FETs) $t_g \approx 0.1 \mu$s. For electrons in gate-defined  GaAs  QDs $t_g \approx 1.5 \mu$s, while for InAs QDs encoded in square fin FETs $t_g\approx 4.3 \mu$s.
 
The fidelity of the CPHASE gate is affected by both the photonic loss of the resonator and the intrinsic decoherence mechanisms. The former is independent of the qubit system and has been discussed at length in~\cite{Royer_Quantum.2017Fast,Harvey_Phys.Rev.B.2018Coupling}
. Therefore, here we only give some estimates based on their findings. The photon-induced dephasing rate is $\Gamma_{ph}\sim2\kappa\left(\frac{g_{z}}{2\Delta}\right)^2$, while the gate infidelity scales as \cite{Royer_Quantum.2017Fast}
\begin{align}
    1-\mathcal{F}_{ph}\sim \Gamma_{ph} t_g\approx\left(Q\frac{eE_c\lambda_0}{\omega_0}\frac{\lambda_0}{\lambda_{SO}}\right)^{-1}\sim\frac{1}{\sqrt{n_{ph}}}\,,
\end{align}
where in the last term we assumed, for simplicity, linear driving with $R_{0,y}/\lambda_{SO}=0.25$ and $n=1$ (which corresponds to the fastest gate). Therefore, the infidelity due to photon leakage $1-\mathcal{F}_{ph}\sim10^{-2}$|$10^{-3}$ for the QD systems discussed in the previous section. The above estimates clearly favor hole-spin implementations: the SOI is strong and can be manipulated by external electrical fields. We note that fidelity can be extended, for example, by squeezing the resonator state \cite{Didier_Phys.Rev.Lett..2015Fast}.
However, a complete description of gate operation is beyond the scope of this work and will be left for future studies.

\section{Geometrical origin of the coupling}
\label{Sec:Geometrical}

\begin{table*}[t]
\begin{ruledtabular}
\begin{tabular}{ccccccc}
   System &  $\omega_0/2\pi$ [GHz]  & $E_c$  [$V m^{-1}$] & $R_c/l_{SO}$ & $g_z/ (2 \pi)$ [MHz] & $n_{ph}$ & $t_g$ [ns]\\
   \hline
Hole-spins in Ge/SiGe heterostructures
\cite{Bosco_Phys.Rev.B.2021Squeezed}
& $10^{2}$ & $1.6 \times 10^{2}$  & $4\times 10^{-3}$  &  $10$ & $10^5$ & $50$ \\
 %
Hole spins in Ge/Si core/shell NW\cite{Kloeffel_Phys.Rev.B.2018Direct}
& $4\times 10^2$ &$3.2 \times 10^2$  & $10^{-3}$ & $2.4$  & $10^4$ & $200$   \\
%
Hole spins in Si square fin FET \cite{Bosco_PRXQuantum.2021Hole}
& $2\times 10^{2}$ & $4\times 10^2$  & $2\times 10^{-2}$ & $5$ & $10^5$ & $100$\\ 
Electron spins in GaAs 2DEG 
\cite{Li_Phys.Rev.B.1996Effectivemass}& $1.1 \times 10^2$ & $1.6 \times 10^2$  & $1.8 \times 10^{-3}$ & $4$  & $10^5$ & $100$\\
\end{tabular}
\end{ruledtabular}
\caption{
\label{estimates1}
Estimates for the longitudinal spin-photon coupling ($g_z$), the average number of photons in the resonator ($n_{ph}$), and the two-qubit CPHASE gate time ($t_g$) for various semiconducting QDs. We assumed $\lambda_0/\lambda_{SO} = 0.2$ ($\lambda_0/\lambda_{SO} = 0.1$) for the holes (electrons).  The cavity frequency is $\omega_c=\Omega=2\pi\times5$ GHz with a quality factor $Q=10^5$. For all systems, we assume $R_0/\lambda_{SO} =0.25$.}
\end{table*}

To elucidate the geometrical underpinnings of the Floquet qubit coupled to the resonator, here we utilize adiabatic perturbation theory to describe the dynamics of a QD in the presence of the microwave resonator in the limit $\Omega\ll\omega_0$.  As already anticipated in Sec.~\ref{Sec:Decoherence},  a unitary  transformation $\mathcal{U}[{\bs R}(t)]={\rm e}^{-i{\bs R}(t)\cdot {\bs p}}$  on $H_{\rm tot}(t)$  renders the confining potentials static, at the expense of two novel terms: 
  \begin{equation}
          \widetilde{H}_{\rm tot}(t)=H_{\rm tot}(0)-\dot{\bs R}\cdot{\bs p}+ e{\bs E}_c\cdot{\bs R}(a^\dagger+a)\,. 
          \label{full}
  \end{equation}
The second term ($\propto \dot{\bs R}$) quantifies a Galilean boost associated with uniform translations, while the third term ($\propto{\bs R}$) represents a time-dependent displacement in the photonic field.  However, it is easy to check that the latter does not affect the dynamics in the leading order in velocity $\dot{\bs R}$ and can be ignored (it can be gauged in the leading order). In the absence of the resonator and time-dependent electrical fields, each electron or hole QD displays a spectrum of degenerate Kramer doublets, as discussed in Sec.~\ref{Sec:FloquetPhotonInteractions}.

We start by treating the term $\propto {\dot{\bs R}}(t)$  in perturbation theory with respect to the static (and decoupled from the photons) QD Hamiltonian $H_{el}(0)$. That is, we perform a (time-dependent) Schrieffer-Wolff transformation of the total Hamiltonian, $\mathcal{U}(t)=e^{S}=1+S+S^2/2+\dots$, such that $[S,H_{el}(0)]+(1-\mathcal{P}_0){\dot{\bs R}}\cdot{\bs p}=0$, where $\mathcal{P}_0$ is a projector onto the lowest degenerate subspace. Reinstating the electron-photon coupling, the resulting Hamiltonian becomes 
\begin{align}
H_{s-p}(t)=\frac{\dot{R}_\alpha}{\lambda_{SO}}\left(\mathcal{A}_\alpha-\frac{R_{c,\beta}}{\lambda_{SO}}\mathcal{F}_{\alpha\beta}(a^\dagger+a)\right)+\omega_c a^\dagger a\,,\nonumber
\end{align}
where $\mathcal{A}_\alpha\equiv\mathcal{P}_0p_\alpha\mathcal{P}_0=m_\alpha$ and  $\mathcal{F}_{\alpha\beta}=i[\mathcal{A}_\alpha,\mathcal{A}_\beta]=2\sigma_z\epsilon_{\alpha\beta z}/\lambda_{SO}^2$ with $\alpha=x,y$ and $\epsilon_{\alpha\beta\gamma}$ the Levi-Civita symbol,  are the non-Abelian Berry connection and curvature, respectively, associated with the electron or hole spin confined in the SO coupled QD \cite{Snizhko_Phys.Rev.B.2019NonAbelian}.
Since $S\propto\dot{\bs R}$, the neglected terms are $\propto(\dot{\bs R})^2,\ddot{\bs R}$. Therefore, in this description, the isolated dynamics is dictated by the Berry connection, whereas the coupling to the resonator is mediated by the non-Abelian Berry curvature \cite{Snizhko_Phys.Rev.B.2019NonAbelian}.
When the driving is periodic, ${\bs R}(t+T)={\bs R}(t)$, it is straightforward to show that the above Hamiltonian reproduces both the linearly polarized (Eq.~\eqref{linear}) and circularly polarized drive (Eqs.~\eqref{long_circ},\eqref{trans_circ}) results.

The adiabatic framework allows us to extend to cases where the QD confinement is arbitrary, not only harmonic. Then, the time-dependent electric field ${\bs E}(t+T)={\bs E}(t)$ not only shifts the center of the QD, but also modifies its shape. In the case when the minimum instantaneous gap (during
the cyclic evolution) between the qubit subspace and the rest of the energy subspaces is
much larger than the driving frequency, following~\cite{Snizhko_Phys.Rev.B.2019NonAbelian}
, we use a unitary transformation $\mathcal{U}\equiv\mathcal{U}[{\bs E}(t)]$ which diagonalises the Hamiltonian in the instantaneous basis including the drive $H_{el}(t)$, to find:
\begin{align}
H_{s-p}(t)=\dot{E}_\alpha\left[\mathcal{A}^E_\alpha+E_{c,\beta}\mathcal{F}^E_{\alpha\beta}(a^\dagger+a)\right]+\omega_c a^\dagger a\,,
\end{align}
where now $\mathcal{A}^E_\alpha=i\mathcal{U}^\dagger\partial_{E_\alpha}\mathcal{U}$ and $\mathcal{F}^E_{\alpha\beta}=\partial_\alpha\mathcal{A}^E_{\beta}-\partial_\beta\mathcal{A}^E_{\alpha}+i[\mathcal{A}^E_{\alpha},\mathcal{A}^E_{\beta}]$ are, respectively, the Berry connection and curvature pertaining to the changes in the applied electrical field. The evolution operator is determined by $H_s^0(t)=\dot{E}_\alpha\mathcal{A}_\alpha^E$, which can be formally written in the same form as in Eq.~\eqref{evolution}, with $j=0,1$. Finally, the spin-photon coupling in the interaction picture becomes :
\begin{align}
    H_{s-p}= E_{c,\beta}\dot{E}_\alpha\left(m^z_{\alpha\beta}\tau_z^F+m^+_{\alpha\beta}e^{i\epsilon_qt}\tau_+^F+{\rm h. c. }\right)(a^\dagger+a)\nonumber\,,
\end{align}
where $m^z_{\alpha\beta}(t)=(\langle\psi_1(t)|\mathcal{F}_{\alpha\beta}^E|\psi_1(t)\rangle-\langle\psi_0(t)|\mathcal{F}_{\alpha\beta}^E|\psi_0(t)\rangle)$ and $m^+_{\alpha\beta}(t)=\langle\psi_0(t)|\mathcal{F}_{\alpha\beta}^E|\psi_1(t)\rangle=[m^-_{\alpha\beta}(t)]^*$ are periodic functions $\propto\Omega$ (because of $\dot{E}_\alpha$), and $\epsilon_q\propto\Omega$ is the quasienergy. Therefore, in perfect analogy with harmonic confinement, by tuning the cavity in resonance with $\Omega$ ($\epsilon_q$) it is possible to activate the term $\propto\tau_z^F$ ($\propto\tau_+^F$) which corresponds to a longitudinal (transverse) spin-photon coupling.

\section{Conclusions and outlook}
\label{Sec:Outlook}

In this work, we studied the interaction between spins (electron or hole) in QDs that are periodically driven by electrical fields and the photons in a microwave resonator. Using Floquet theory, we showed that spin-orbit interactions induce a time-dependent magnetic field that acts within the ground-state Kramer doublets and generates spin-photon interactions in the absence of any applied magnetic fields. We found both transverse and longitudinal coupling of the Floquet spin-qubit to the resonator, the latter being a novel contribution having no analogue for static Zeeman-split spin qubits. We demonstrated that such longitudinal interactions facilitate both the readout of the qubit and the implementation of a CPHASE gate in the case of two remote QDs coupled to the same cavity. Finally, using adiabatic perturbation theory, we have uncovered the geometrical origin of these dynamical effects and demonstrated their generality to arbitrary QD confinements. A summary of our results is presented in Table \ref{estimates1}, for both hole and electron spins in semiconducting nanostructures. 

There are several possible future directions.  First, it is straightforward to generalize to more driven spin-qubits because the QDs do not interact directly with each other. A more nontrivial extension, in particular for hole-spins in QDs, is to determine microscopically the effect of the electrical fields on both the SOI and the dynamics \cite{Wang_npjQuantumInf.2021Optimal}.
Furthermore, it would be beneficial to evaluate phonon-induced decoherence for the driven hole-spin qubits for more realistic confining potentials and identify sweet spots for their operation. Further down the road, it would be interesting to extract the statistics of the electromagnetic field emitted because of the driving and establish the imprints of the spin-qubit geometry of states onto the photons \cite{Haroche_.2006Exploring,Westig_Phys.Rev.Lett..2017Emission}.

In conclusion, the Floquet spin-qubit proposed in this work operates well within the current experimental capabilities, and we expect it to open up new possibilities for future studies on driven spin-qubits in the absence of externally applied magnetic fields. 


\begin{acknowledgments}
We thank Lukasz Cywinski for useful discussions.  This work was supported by the Foundation for Polish Science through the International Research Agendas (IRA) program co-financed by the European Union within Smart Growth Programme, and by the National Science Centre (Poland) OPUS 2021/41/B/ST3/04475. 
\end{acknowledgments}



\revappendix  
\begin{widetext}
\section{Derivation of the projected Hamiltonian for harmonically confined  QD}\label{projectesdintHamiltonian}

\subsection*{Static QD ($\mathbf{E}(t)=0$) subjected to an in-plane magnetic field}\label{staticcase}

When the confinement of the QDs is harmonic, and the QD is subject to an in-plane external magnetic field, such that it only causes a Zeeman coupling, we can establish the following relationships between various matrix elements (setting $\hbar=1$) \cite{San-Jose_Phys.Rev.B.2008Geometric}:
\begin{align}
\langle\psi_j|[{\bs p},H_{el}]|\psi_{j'}\rangle&=(\epsilon_{j'}-\epsilon_{j})\langle\psi_ j|{\bs p}|\psi_{j'}\rangle\equiv -i m\omega_0^2\langle\psi_j|{\bs r}|\psi_{j'}\rangle
\end{align}
and
\begin{align}
\langle \psi_j|[{\bs r}, H_{el}]|\psi_{j'}\rangle&=(\epsilon_{j'}-\epsilon_{j})\langle \psi_j| {\bs r}|\psi_{j'}\rangle\equiv\frac{i}{m}\langle \psi_j|{\bs P}|\psi_{j'}\rangle\,.    
\end{align}
Above, ${\bs P}={\bs p}+{\bs m}/\lambda_{SO}$, ${\bs m}=\lambda_{SO}{\bs \lambda}_{SO}^{-1}{\bs \sigma}$, ${\bs\lambda}_{SO}^{-1} = \left[\begin{smallmatrix}
    -\beta &\alpha \\
    -\alpha & \beta 
\end{smallmatrix}\right]$, $\lambda_{SO}^{-1}=||{\bs \lambda}_{SO}^{-1}||/\sqrt{2}=m\sqrt{\alpha^2 +\beta^2 }$,
 and $|\psi_j\rangle $ are the eigenstates of $H_{el}$. 
That in turn implies
\begin{align}
-\langle \psi_j| {\bs p}|\psi_{j'}\rangle&=\frac{1}{\lambda_{SO}}\frac{\omega_0^2}{\omega_0^2-(\epsilon_{j}-\epsilon_{j'})^2}\langle \psi_{j'}|{\bs m}|\psi_{j'}\rangle\,,\\
-\langle \psi_j| {\bs r}|\psi_{j'}\rangle&=-\frac{i}{m\lambda_{SO}}\frac{\epsilon_{j}-\epsilon_{j'}}{\omega_0^2-(\epsilon_{j}-\epsilon_{j'})^2}\langle \psi_j|{\bs m}|\psi_{j'}\rangle\,.
\end{align}
Hence, the momentum and position operators, respectively, projected onto the two lowest states $\{|\psi_0\rangle,|\psi_1\rangle\}$ become:
\begin{align}
    \tilde{\bs p}&\equiv\mathcal{P}_0{\bs p}\mathcal{P}_0=\frac{1}{\lambda_{SO}}\left(\frac{\omega_0^2}{\omega_0^2-\epsilon_{q}^2}({\bs m}_{+}\,\tau_++{\bs m}_{-}\,\tau_-)+\frac{1}{2}{\bs m}_z\tau_z\right)\,,\\   
    \tilde{\bs r}&=\mathcal{P}_0{\bs r}\mathcal{P}_0=\frac{i}{m\lambda_{SO}}\frac{\epsilon_{q}}{\omega_0^2-\epsilon_{q}^2}({\bs m}_{+}\tau_+-{\bs m}_{-}\tau_-)\,,
    \label{relations}
\end{align}
where $\mathcal{P}_0$ is the projector, $\epsilon_q=\epsilon_1-\epsilon_0$,  and 
\begin{align}
    {\bs m}_{+}&=({\bs m}_{-})^*=\langle\psi_1|{\bs m}|\psi_0\rangle\nonumber \,,\\ 
    {\bs m}_z&=\langle\psi_1|{\bs m}|\psi_1\rangle-\langle\psi_0|{\bs m}|\psi_0\rangle\,.
\end{align}
In the above expressions, ${\bs \tau}=(\tau_x,\tau_y,\tau_z)$ are Pauli matrices acting on the subspace defined by the pair $\{|\psi_0\rangle,|\psi_1\rangle\}$. These expressions are valid in the absence of the coupling to the photons and account, as already mentioned, for a Zeeman coupling.  Therefore, the electron-photon Hamiltonian acting in the doublet subspace reads:
\begin{align}
H_{s-p}&=\mathcal{P}_0H_{e-p}\mathcal{P}_0=i\frac{e}{m\lambda_{SO}}\frac{\epsilon_q}{\omega_0^2-\epsilon_q^2}{\bs E}_{c}\cdot({\bs m}_{+}\tau_+-{\bs m}_{-}\tau_-)(a^\dagger+a)\approx i\epsilon_q\frac{{\bs R}_c}{\lambda_{SO}}\cdot({\bs m}_{+}\tau_+-{\bs m}_{-}\tau_-)(a^\dagger+a)\,,
\end{align}
 where in the last line we assumed $\epsilon_q\ll\omega_0$ and defined ${\bs R}_c=\frac{e}{m}\frac{{\bs E}_{c}}{\omega_0^2}$ as the distance the cavity field electrical field shifts the center of the QD. This represents the expression shown in the main text.

\subsection*{Dynamical  case [$\mathbf{E}(t) \neq 0$] in zero applied magnetic fields}\label{dynamiccase}

The relationships for the static case can be extended to the periodically driven system. Starting from the matrix element of an operator $\mathcal{A}$ of the system in the interaction picture becomes: 
\begin{align}
\mathcal{A}^I(t)=\sum_{j,j'}\langle\psi_j(t)|\mathcal{A}|\psi_{j'}(t)\rangle|\psi_j(0)\rangle\langle\psi_{j'}(0)|e^{i(\epsilon_j-\epsilon_{j'})t}\,,
\end{align}
where $\epsilon_j$ is the quasienergy,   $|\psi_j(t)\rangle=|\psi_j(t+T)\rangle$ is the periodic part of the Floquet states $|\Psi_j(t)\rangle=e^{-i\epsilon_jt}|\psi_j(t)\rangle$, and the summation contains all the Floquet states, including the Floquet spin-qubit. Moreover:
\begin{align}
    \langle\psi_j(t)|\mathcal{A}|\psi_{j'}(t)\rangle&=\sum_ke^{ik\Omega  t}\mathcal{A}_{jj'}(k)\nonumber\,,\\
    \mathcal{A}_{jj'}(k)&=\frac{1}{T}\int_0^Tdte^{-ik\Omega  t}\langle\psi_j(t)|\mathcal{A}|\psi_{j'}(t)\rangle\,.
\end{align}

Therefore, in the Fourier space we obtain the following identities: 
\begin{align}
(\epsilon_j-\epsilon_{j'}+k\Omega){\bs p}_{jj'}(k)&=-i m\omega_0^2{\bs r}_{jj'}(k)\nonumber\,,\\
(\epsilon_j-\epsilon_{j'}+k\Omega){\bs r}_{jj'}(k)&=\frac{i}{m}{\bs P}_{jj'}(k)\,.
\end{align}
Finally, we can establish  relationships analogous to the static case:
\begin{align}
{\bs p}_{jj'}(k)&=-\frac{1}{\lambda_{SO}}\frac{\omega_0^2}{\omega_0^2-(\epsilon_j-\epsilon_{j'}+k\Omega)^2}{\bs m}_{jj'}(k)\,,\\
{\bs r}_{jj'}(k)&=\frac{i}{m\lambda_{SO}}\frac{\epsilon_j-\epsilon_{j'}+k\Omega}{\omega_0^2-(\epsilon_j-\epsilon_{j'}+k\Omega)^2}{\bs m}_{jj'}(k)\,,
\end{align}
where ${\bs m}_{jj'}(k)$ is the $k$-th Fourier component of $\langle\psi_j(t)|{\bs m}|\psi_{j'}(t)\rangle$. These identities can be used to express the electron-photon Hamiltonian in the Floquet basis (hence, in the interaction picture) as follows: 
\begin{align}
    H_{e-p}^I(t)&=i\frac{e}{m\lambda_{SO}}\sum_{j,j'}\sum_ke^{i(\epsilon_j-\epsilon_{j'}+k\Omega)t}\frac{\epsilon_j-\epsilon_{j'}+k\Omega}{\omega_0^2-(\epsilon_j-\epsilon_{j'}+k\Omega)^2}{\bs E}_c\cdot{\bs m}_{jj'}(k)|\psi_j(0)\rangle\langle\psi_{j'}(0)|(a^\dagger e^{i\omega_ct}+ae^{-i\omega_ct})\,.
\end{align} 
We can further simplify this expression when the driving is adiabatic. Specifically,  when \begin{align}
    \Omega\ll{\rm min}|\bar{\epsilon}_j-\bar{\epsilon}_{j'}|\,,
\end{align}
for levels levels $j$ and $j'$ that do not belong to the same Kramers doublet in the absence of driving and, $\bar{\epsilon}_j=(1/T)\int_0^T dt\langle\psi_j(t)|H_{el}(t)|\psi_j(t)\rangle$ is the average energy. Therefore, we define: 
\begin{align}    \tau_x^F&=|\psi_0(0)\rangle\langle\psi_1(0)|+|\psi_1(0)\rangle\langle\psi_0(0)|\nn\,,\\
    \tau_y^F&=i(|\psi_0(0)\rangle\langle\psi_1(0)|-|\psi_1(0)\rangle\langle\psi_0(0)|)\,,\\
    \tau_z^F&=|\psi_1(0)\rangle\langle\psi_1(0)|-|\psi_0(0)\rangle\langle\psi_0(0)|\nn\,,
\end{align}
which constitutes the dynamical analogue of the static Zeeman-split spin-qubit levels. Projecting the interaction Hamiltonian onto this subspace we find:
\begin{align}
    H_{s-p}^I(t)&\approx\frac{ie\Omega}{m\omega_0^2 \lambda_{SO}}\sum_kke^{ik\Omega t}{\bs E}_c\cdot[{\bs m}_{11}(k)-{\bs m}_{00}(k)]\frac{\tau_z^F}{2}(a^\dagger e^{i\omega_ct}+{\rm h. c.})\nonumber\\
    &+\frac{ie}{m\omega_0^2\lambda_{SO}}\sum_k(\epsilon_q+k\Omega){\bs E}_c\cdot\left[e^{i(\epsilon_q+k\Omega)t}{\bs m}_{10}(k)\tau_+^F-e^{-i(\epsilon_q+k\Omega)t}{\bs m}_{01}(-k)\tau_-^F\right](a^\dagger e^{i\omega_ct}+{\rm h. c.})\nonumber\\
    &=\left(\underbrace{\frac{{\bs R}_c}{2 \lambda_{SO}}\cdot\frac{d}{dt}{\bs m}_{z}(t)}_{\displaystyle  g_z(t)}\tau_z^F+\underbrace{i\frac{{\bs R}_c}{\lambda_{SO}}\cdot e^{i\epsilon_q t}\left(\epsilon_q-i\frac{d}{dt}\right){\bs m}_{+}(t)}_{\displaystyle g_{+}(t)}\tau_+^F+{\rm h. c. }\right)(a^\dagger e^{i\omega_ct}+{\rm h. c.})\,,
\end{align}
where $\epsilon_q=\epsilon_1-\epsilon_0$, while we defined  ${\bs m}_{\pm}(t)=\langle\psi_ {1,0}(t)|{\bs m}|\psi_{0,1}(t)\rangle$ and ${\bs m}_z(t)=\langle\psi_1(t)|{\bs m}|\psi_{1}(t)\rangle-\langle\psi_0(t)|{\bs m}|\psi_{0}(t)\rangle$. 


\subsection*{Spin-photon coupling for linearly polarized electrical driving}\label{linearpath}

In this subsection, we provide additional information specifically for the scenario in which the QD is subject to linear driving. We assume that the QD is driven along the $y$-direction in the semiconductor, or
\begin{align}
    {\bs R}(t)={\bs e}_yR_{0,y}\sin(\Omega t)\,,
\end{align}
where $R_{0,y}$ is the amplitude of the drive.  It is instructive to switch to a frame moving with the QD, which is achieved by applying a unitary transformation $\mathcal{U}[{\bs R}(t)]=e^{-i{\bs p}\cdot{\bs R}(t)}$ to the total Hamiltonian. The Hamiltonian for the lowest doublet is then transformed into
\begin{align}
    H_{s}(t)=\frac{\dot{\bs R}(t)}{\lambda_{SO}}\cdot{\bs m}\,,
\end{align}
where we only kept the lowest order effects in the SOI (otherwise, ${\bs \lambda}_{SO}^{-1}{\bs \sigma}$ needs to be substituted by $\mathcal{P}_0{\bs \lambda}_{SO}^{-1}{\bs \sigma}\mathcal{P}_0$). To streamline the discussion, we make the assumption that only Rashba SOI is present. Consequently, we can express the above equation as follows:
\begin{align}
    H_{s}(t)=-\frac{\Omega R_{0,y}}{\lambda_{SO}}\sigma_x\cos(\Omega t)\,.
\end{align}
The corresponding Floquet quasienergies and states are, respectively, $\epsilon_{1,0}=0$ and
\begin{align}
    |0,1(t)\rangle=\frac{1}{\sqrt{2}}(|\uparrow\rangle\pm|\downarrow\rangle)\rangle 
    \exp\left[\pm i\frac{ R_{0,y}\sin(\Omega t)}{\lambda_{SO}} \right]
    =\frac{1}{\sqrt{2}}(|\uparrow\rangle\pm|\downarrow\rangle)\sum_{p\in\mathcal{Z}}e^{\displaystyle\pm ip\Omega t}J_p(R_{0,y}/\lambda_{SO})\,,
\end{align}
where $J_p(x)$ is the Bessel function of the second kind. Since these states only have global phases $g_{z}(t)\equiv0$, the second term generates the entire spin-photon Hamiltonian:
\begin{align}
    H_{s-p}(t)&=2\Omega\frac{R_{c,x} R_{0,y}}{\lambda_{SO}^2}\cos(\Omega t)
    \exp \left[ \pm 2i\frac{R_{0,y}\sin(\Omega t)}{\lambda_{SO}} \right]
    \tau_+^F(a^\dagger e^{i\omega_ct}+a e^{-i\omega_ct})+{\rm h. c.}\,,
\end{align}
where $R_{c,x}=eE_{c,x}/m\omega_0^2$ quantifies the displacement of the QD by the cavity electric field. Assuming that the driving frequency is tuned close to the cavity frequency, $\Omega\sim\omega_c$, we can retain in the above expression only the terms that slowly oscillate on the scales $\Omega$ and $\omega_c$. These terms read:
\begin{align}
    H_{s-p}&\approx2\Omega\frac{R_{c,x} R_{0,y}}{\lambda_{SO}^2}\tau_x^F(a^\dagger+a)\,,
\end{align}
as depicted in Eq.~\eqref{linear} in the main text.


\section{Details on the input-output approach for detection}\label{inputoutputdetails}

The equation of motion for the photon field reads
\begin{align}
    \dot{a}(t)&=-i\omega_ca(t)+ieE_{c,\alpha}r_\alpha(t)-\frac{\kappa}{2}a(t)-\sqrt{\kappa}a_{in}(t)\,,
\end{align}
where $r_\alpha(t)$ is the position operator in the Heisenberg picture. In leading order to the coupling to the photons, we can express it as:
\begin{align}
r_\alpha(t)&\approx r^I_\alpha(t)+ieE_{c,\alpha}\int_{-\infty}^td\tau(a^\dagger+a)(\tau)[r_\alpha^I(\tau),r_\alpha^I(t)]\,,
\end{align}
where $r_\alpha^I(t)$ is the position operator in the interaction picture. We assume that the cavity is driven continuously with $\epsilon(t)=\epsilon_d\exp(i\omega_d t)$, with $\omega_d$ being the driving frequency. Then, the input field can be characterized by its mean $\alpha_{in}(t)=\langle a_{in}(t)\rangle$, where $\alpha_{in}(t)=\epsilon(t)/\sqrt{\kappa}$. Fluctuations around the input value will determine the magnitude of the noise in the detection signal. On top of that, the following relation holds between the input, output, and the cavity field:
\begin{align}
    a_{out}(t)=a_{in}(t)+\sqrt{\kappa}a(t)\,,
\end{align}
which also holds for the averages $\alpha_{out}(t)=\alpha_{in}(t)+\sqrt{\kappa}\alpha(t)$. Thus, the equation governing the average cavity field is as follows: 
\begin{align}
\dot{\alpha}(t)\approx-i\omega_c\alpha(t)+iE_{c,\alpha}\langle r^I_\alpha(t)\rangle-i\alpha(t) \int_{-\infty}^\infty d\tau e^{-i\omega_c\tau}\chi_{\alpha}(t-\tau,t)-\frac{\kappa}{2}\alpha(t)-\sqrt{\kappa}\alpha_{in}(t)\,,
\end{align}
where
\begin{align}
\chi_{\alpha}(t',t)&=-i\theta(t-t')E_{c,\alpha}^2\langle[r_\alpha^I(t'),r_\alpha^I(t)]\rangle\,,
\end{align}
is the time-dependent susceptibility of the isolated system, and $\langle\dots\rangle$ represents the trace over the stationary density matrix of the electrons in the absence of the coupling to photons. For periodic driving in the stationary regime, the susceptibility obeys $\chi_{\alpha}(t',t)=\chi_{\alpha}(t'+T,t+T)$. Hence, it can be formally written as:
\begin{align}
    \chi_{\alpha}(t-\tau,t)=\sum_k\frac{1}{2\pi}\int d\omega e^{-ik\Omega t-i\omega\tau}\chi_{\alpha}^k(\omega)\,,
\end{align}
which in turn allows to write the Fourier transformed equation of motion:
\begin{align}
    -i\omega \alpha(\omega)&=-i\omega_c\alpha(\omega)+\sum_q\chi_{\alpha}^q(\omega)\alpha(\omega-q\Omega)-\frac{\kappa}{2}\alpha(\omega)+iE_{c,\alpha}\langle r^I_\alpha(\omega)\rangle-\sqrt{\kappa}\alpha_{in}(\omega)\,.
\end{align}
Assuming that the density matrix in the interaction picture is $\rho_S=\sum_{\tau}\rho_{\tau}|\psi_{\tau}(0)\rangle\langle\psi_{\tau}(0)|$, with $\rho_{\tau}$ being the weight of the Floquet state $\tau=\pm$, we find for the $q=0$ Fourier component: 
\begin{align}
\chi_{\alpha}^0(\omega)&=E_{c,\alpha}^2\sum_{\tau,\tau'}\sum_{k}\rho_{\tau}\left[\frac{r^{\alpha}_{\tau \tau'}(-k)r^{\alpha}_{\tau'\tau}(k)}{\omega+\epsilon_{\tau}-\epsilon_{\tau'}+k\Omega+i\delta}-\frac{r^{\alpha}_{\tau\tau'}(k)r^{\alpha}_{\tau'\tau }(-k)}{\omega-\epsilon_{\tau}+\epsilon_{\tau'}+k\Omega+i\delta}\right]\nonumber\\
    &=E_{c,\alpha}^2\sum_{\tau}\sum_{k}(\rho_{\tau}-\rho_{\bar{\tau}})\frac{r^{\alpha}_{\tau\bar{\tau}}(-k)r^{\alpha}_{\bar{\tau}\tau}(k)}{\omega+\epsilon_{\tau}-\epsilon_{\bar{\tau}}+k\Omega+i\delta}\nonumber\\
    &=\frac{E_{c,\alpha}^2}{( m\omega_0^2)^2}\sum_{\tau}\sum_{k}(\rho_{\tau}-\rho_{\bar{\tau}})\frac{(\epsilon_\tau-\epsilon_{\bar{\tau}}+k\Omega)^2}{\omega+\epsilon_{\tau}-\epsilon_{\bar{\tau}}+k\Omega+i\delta}\mathcal{M}_{\tau\bar{\tau}}^\alpha(-k)\mathcal{M}_{\bar{\tau}\tau}^\alpha(k)\,,
\end{align}
where $\delta$ quantifies the linewidth of the Floquet levels.

Let us evaluate the response when the QD is driven on a circular trajectory.  Then, using the Floquet states in Eq.~\eqref{Floquet_states} in the main text, we find
    \begin{align}
    \sigma_{01,y}(k)=\frac{i}{T}\int_0^T  dte^{-ik\Omega t}[e^{2i\Omega t}\sin^2(\theta/2)+\cos^2(\theta/2)]=i [\sin^2(\theta/2)\delta_{k,2}+\cos^2(\theta/2)\delta_{k0}]=[\sigma_{10,y}(-k)]^*\nonumber\,,\\
    \sigma_{01,x}(k)=\frac{1}{T}\int_0^T  dte^{-ik\Omega t}[e^{2i\Omega t}\sin^2(\theta/2)-\cos^2(\theta/2)]= \sin^2(\theta/2)\delta_{k,2}-\cos^2(\theta/2)\delta_{k0}=[\sigma_{10,x}(-k)]^*\,,
\end{align}
which can be substituted in the expression for the susceptibility to give
\begin{align}  &\chi_{\alpha}^0(\omega)=\left(\frac{R_{c,\alpha}}{\lambda_{SO}}\right)^2\sum_{\tau}\sum_{k}(\rho_{\tau}-\rho_{\bar{\tau}})\frac{(\epsilon_\tau-\epsilon_{\bar{\tau}}+k\Omega)^2}{\omega+\epsilon_{\tau}-\epsilon_{\bar{\tau}}+k\Omega+i\delta}[\sin^4(\theta/2)\delta_{k,-2}+\cos^4(\theta/2)\delta_{k,0}]\nonumber\\
    &=\left(\frac{R_{c,\alpha}}{\lambda_{SO}}\right)^2\sum_{\tau}(\rho_{\tau}-\rho_{\bar{\tau}})\left[\frac{(\epsilon_\tau-\epsilon_{\bar{\tau}}-2\Omega)^2}{\omega+\epsilon_{\tau}-\epsilon_{\bar{\tau}}-2\Omega+i\delta}\sin^4(\theta/2)+\frac{(\epsilon_\tau-\epsilon_{\bar{\tau}})^2}{\omega+\epsilon_{\tau}-\epsilon_{\bar{\tau}}+i\delta}\cos^4(\theta/2)\right]\nonumber\\
    &\approx\frac{(\Omega\gamma_B)^2}{4(\gamma_B+1)^2}\left(\frac{R_{c,\alpha}}{\lambda_{SO}}\right)^2(\rho_{+}-\rho_{-})\left[\frac{(\gamma_B-2)^2}{\omega+(\gamma_B-2)\Omega+i\delta}-\frac{(\gamma_B+2)^2}{\omega-(\gamma_B+2)\Omega+i\delta}+\frac{\gamma_B^2}{\omega-\gamma_B\Omega+i\delta}\right]\,.
\end{align}
We see that when $\omega\sim\gamma_B\Omega$, this simplifies the result obtained in the main text, while the response vanishes in the limit $\gamma_B\rightarrow0$, in accordance with the findings for linearly polarized driving.

\section{Floquet-Born-Markov approach to dissipation}\label{Floquet-Born-Markov}

Here we briefly summarize the derivation of the rate equation for the Floquet spin-qubit in the presence of bosonic environments. Assuming that the relevant dynamics occurs within the originally degenerate subspace, the Floquet qubit-bath interaction Hamiltonian reads (in the interaction picture):
\begin{align}
    H_{s-b}^I(t)&=\sum_{\alpha=x,y} \; \underbrace{
    \sum_{k \in \mathbb{Z}}
    \left[\widetilde{r}^z_{\alpha}(k)\tau_z^F +\widetilde{r}_{\alpha}^+(k)\tau_+^Fe^{+i\epsilon_qt}+\widetilde{r}_{\alpha}^-(k)\tau_-^Fe^{-i\epsilon_qt}\right]e^{ik\Omega t}}_{\displaystyle A_{\alpha}(t)}\mathcal{E}_\alpha(t)\,,
\label{qubit-boson}
\end{align} 
where $\widetilde{r}_\alpha^{z,+,-}(k)=r_\alpha^{z,+,-}(k)/\lambda_0$ and  we only retained the terms that do not act as identity in this subspace. In this case, we assume the entire population of the QD is localized in this subspace and, therefore, describe the evolution of the Floquet spin-qubit by its density matrix (in the Born approximation):
\begin{align}
    \dot{\rho}_S(t)&=-\int_{t_0}^tdt'{\rm Tr}_B\left\{H_{s-b}^I(t),[H_{s-b}^I(t'),\rho_S(t)\otimes\rho_B]\right\}=-\sum_{\alpha, \beta}\int_{t_0}^tdt'J_{\alpha\beta}(t-t')[A_{\alpha}(t),A_{\beta}(t')\rho_S(t')]+{\rm h. c. }
\end{align}
where 
\begin{equation}
    J_{\alpha\beta}(t-t')={\rm Tr}_B[\mathcal{E}_\alpha(t) \mathcal{E}_\beta(t')\rho_B]\delta_{\alpha\beta}\equiv J_{\alpha}(t-t')\,,
\end{equation}
being the bath correlation function and $\rho_B$ is the density matrix of the bath (for simplicity, we have assumed the cross correlations between different components vanish). By also applying the Markov approximation, which assumes that $\rho_S(t')\approx\rho_S(t)$ on the right-hand side, we can determine the evolution of the density matrix as follows (taking $t_0 \rightarrow -\infty$):
\begin{align}
   &\dot{\rho}_S(t)=-\int_{-\infty}^\infty d\omega\int_{-\infty}^tdt'\sum_\alpha J_{\alpha}(\omega)e^{i\omega(t-t')}\sum_{k,k'}e^{ik\Omega t}e^{ik'\Omega t'}\nonumber\\
   &\times\bigg[
    (\widetilde{r}^z_{\alpha}(k)\tau_z^F +\widetilde{r}_{\alpha}^+(k)\tau_+^Fe^{i\epsilon_qt}+\widetilde{r}_{\alpha}^-(k)\tau_-^Fe^{-i\epsilon_qt})(\widetilde{r}^z_{\alpha}(k')\tau_z^F +\widetilde{r}_{\alpha}^+(k')\tau_+^Fe^{i\epsilon_qt'}+\widetilde{r}_{\alpha}^-(k')\tau_-^Fe^{-i\epsilon_qt'})\rho_S(t)\nonumber\\
    &-(\widetilde{r}^z_{\alpha}(k')\tau_z^F +\widetilde{r}_{\alpha}^+(k')\tau_+^Fe^{i\epsilon_qt'}+\widetilde{r}_{\alpha}^-(k')\tau_-^Fe^{-i\epsilon_qt'})\rho_S(t)(\widetilde{r}^z_{\alpha}(k)\tau_z^F +\widetilde{r}_{\alpha}^+(k)\tau_+^Fe^{i\epsilon_qt}+\widetilde{r}_{\alpha}^-(k)\tau_-^Fe^{-i\epsilon_qt})\bigg]+{\rm h. c. }\\
    &=-i\int_{-\infty}^\infty d\omega\sum_\alpha J_{\alpha}(\omega)\sum_{k,k'}e^{i(k+k')\Omega t}\nonumber\\
   &\times\bigg[
    (\widetilde{r}^z_{\alpha}(k)\tau_z^F +\widetilde{r}_{\alpha}^+(k)\tau_+^Fe^{i\epsilon_qt}+\widetilde{r}_{\alpha}^-(k)\tau_-^Fe^{-i\epsilon_qt})\left(\frac{\widetilde{r}^z_{\alpha}(k')}{\omega-k' \Omega+i\eta}\tau_z^F +\frac{\widetilde{r}_{\alpha}^+(k' )e^{i\epsilon_qt}}{\omega-\epsilon_q-k' \Omega+i\eta}\tau_+^F+\frac{\widetilde{r}_{\alpha}^-(k' )e^{-i\epsilon_qt}}{\omega+\epsilon_q-k' \Omega+i\eta}\tau_-^F\right)\rho_S(t)\nonumber\\
    &-\left(\frac{\widetilde{r}^z_{\alpha}(k' )}{\omega-k' \Omega+i\eta}\tau_z^F +\frac{\widetilde{r}_{\alpha}^+(k' )e^{i\epsilon_qt}}{\omega-\epsilon_q-k' \Omega+i\eta}\tau_+^F+\frac{\widetilde{r}_{\alpha}^-(k' )e^{-i\epsilon_qt}}{\omega+\epsilon_q-k' \Omega+i\eta}\tau_-^F\right)\rho_S(t)(\widetilde{r}^z_{\alpha}(k)\tau_z^F +\widetilde{r}_{\alpha}^+(k)\tau_+^Fe^{i\epsilon_qt}+\widetilde{r}_{\alpha}^-(k)\tau_-^Fe^{-i\epsilon_qt})\bigg]+{\rm h. c. }\nonumber\,.
\end{align}
Finally, in order to obtain a Lindblad-type density matrix evolution, we perform the secular approximation, which means keeping in the above expression only the terms that oscillate slow compared to the dynamics of the qubit. Assuming non-zero $\epsilon_q\neq0$ ($\epsilon_q=0$ needs to be treated separately) and recalling that $-\Omega/2<\epsilon_q<\Omega/2$, implies that $k' =-k$ in the above expressions. That in turn gives
\begin{align}
\dot{\rho}_S(t)&=-i\int d\omega\sum_\alpha J_{\alpha}(\omega)\sum_{k}\bigg[
    \left(\frac{|\widetilde{r}^z_{\alpha}(k)|^2}{\omega+k\Omega+i\eta}\tau_z^F\tau_z^F +\frac{|\widetilde{r}_{\alpha}^-(k)|^2}{\omega-\epsilon_q+k\Omega+i\eta}\tau_-^F\tau_+^F+\frac{|\widetilde{r}_{\alpha}^+(k)|^2}{\omega+\epsilon_q+k\Omega+i\eta}\tau_+^F\tau_-^F\right)\rho_S(t)\nonumber\\
    &-\left(\frac{|\widetilde{r}^z_{\alpha}(k)|^2}{\omega+k\Omega+i\eta}\tau_z^F\rho_S(t)\,\tau_z^F +\frac{|\widetilde{r}_{\alpha}^-(k)|^2}{\omega-\epsilon_q+k\Omega+i\eta}\tau_+^F\rho_S(t)\,\tau_-^F+\frac{|\widetilde{r}_{\alpha}^+(k)|^2}{\omega+\epsilon_q+k\Omega+i\eta}\tau_-^F\rho_S(t)\,\tau_+^F\right)\bigg]+{\rm h. c. }\,.
\end{align}
Next, neglecting the real part contribution in the above integrals, which represents the Lamb shift that slightly renormalizes the spectrum, the above expression can be cast in a Lindblad form: \cite{Breuer_.2002Theory}
\begin{align}
    \dot{\rho}_S(t)&=\Gamma_-\left(\tau_-^F\rho_S(t)\,\tau_+^F-\frac{1}{2}\{\tau_+^F\tau_-^F,\rho_S(t)\}\right)+\Gamma_+\left(\tau_+^F\rho_S(t)\,\tau_-^F-\frac{1}{2}\{\tau_-^F\tau_+^F,\rho_S(t)\}\right)\nonumber\\
&+\Gamma_z\left(\tau_z^F\rho_S(t)\,\tau_z^F-\frac{1}{2}\{\tau_z^F\tau_z^F,\rho_S(t)\}\right)\,,
\end{align}
with the rates showed in Eq.~\eqref{rates}.

\section{Evaluation of the energy flow into the environment}
 
Next, we estimate the energy released into the environment by the driving field via the QD. We start by explicitly writing the fluctuating fields in Eq.~\ref{qubit-boson}:
\begin{align}
    \mathcal{E}_\alpha=\int_0^\infty d\omega g_\alpha(\omega)[a_\alpha^\dagger(\omega)+a_\alpha(\omega)]\,,
\end{align}
where $a_\alpha(\omega)$ [$a^\dagger_\alpha(\omega)$], with $\alpha=x,y$, are the annihilation  (creation) operator for the environment bosonic mode of energy $\omega$, which couples to the QD with strength $g_\alpha(\omega)$. The Hamiltonian of the environment is 
\begin{align}
H_{\rm env}=\sum_{\alpha=x,y}\int_0^\infty  d\omega\omega a_\alpha^\dagger(\omega)a_\alpha(\omega)\,.
\end{align}
so that the change in the energy of the bath as a result of the coupling to the QD (or the power) can be written as follows:
\begin{align}
    \hat{P}=\frac{dH_{\rm env}}{dt}=i[H_{\rm env},H_{d-b}]=i\sum_{\alpha=x,y}\int_0^\infty d\omega\omega g_\alpha(\omega)[a_\alpha^\dagger(\omega,t)-a_\alpha(\omega,t)]r_\alpha(t)\,,
\end{align}
where all operators are evolving in the Heisenberg picture. Thus, we are left with determining the average power, $\langle\hat{P}\rangle$, and we do so in the leading order in $g_\alpha(\omega)$. 

The equation of motion (in the Heisenberg) for the bosonic bath operator reads:
\begin{align}
    \dot{a}_\alpha(\omega,t)=-i\omega a_\alpha(\omega,t)+ig_\alpha(\omega)\widetilde{r}_\alpha(t)\,,
\end{align}
which in turn gives for the field itself:
\begin{equation}
    a_\alpha(\omega,t)=a_{\alpha,I}(\omega,t_i)e^{-i\omega(t-t_i)}+i\int_{t_i}^t d\tau g_\alpha(\omega)e^{-i\omega(t-\tau)}\widetilde{r}_\alpha(\tau)\,,
\end{equation}
where $t_i$ is a time in the distant past when the systems did not interact. Next we evaluate $\widetilde{r}_\alpha(\tau)$, which is given by:
\begin{align}
    \widetilde{r}_\alpha(\tau)=U^\dagger_{I,d-b}(\tau,t_i)\widetilde{r}_{\alpha,I}(\tau)U_{I,d-b}(\tau,t_i)\,,
\end{align}
where
\begin{align}
    U_{I,d-b}(\tau,t_i)&=\mathcal{T}e^{-i\int_{t_i}^\tau dt' H_{d-b}^I(t')}\approx 1-i\int_{t_i}^\tau dt' H_{d-b}^I(t')=1-i\int_{t_i}^\tau dt' \widetilde{r}_{\alpha,I}(t')\mathcal{E}_I(t')\nonumber\\
    &=1-i\sum_\alpha\int_{t_i}^\tau dt' \widetilde{r}_{\alpha,I}(t')\int_0^\infty d\omega g_\alpha(\omega)[a_{\alpha,I}^\dagger(\omega,t')+a_{\alpha,I}(\omega,t')]\,.
\end{align}
Therefore, in leading order in $g_\alpha(\omega)$:
\begin{align}
    \widetilde{r}_\alpha(\tau)\approx \widetilde{r}_{\alpha,I}(\tau)-i\sum_\beta\int_{t_i}^\tau dt'\int_0^\infty d\omega g_\alpha(\omega)[\widetilde{r}_{\alpha,I}(\tau),\widetilde{r}_{\beta,I}(t')][a_{\beta,I}^\dagger(\omega,t')+a_{\beta,I}(\omega,t')]\,.
\end{align}
By substituting the above expressions for the bath and position operator in the expression for the power, and retaining only the second order terms in $g_\alpha$, we obtain:
\begin{align}
    \hat{P}(t)
    &=-\sum_\alpha\int_{0}^\infty d\tau\int_0^\infty d\omega\omega g_\alpha^2(\omega)[\widetilde{r}_{\alpha,I}(t-\tau),\widetilde{r}_{\alpha,I}(t)][n_B(\omega)e^{-i\omega\tau}-(1+n_B(\omega))e^{i\omega\tau}]\nonumber\\
    &+\int_0^\infty d\omega\omega g_\alpha^2(\omega)\int_{0}^\infty d\tau (e^{-i\omega\tau}+e^{i\omega\tau})\widetilde{r}_{\alpha,I}(t-\tau)\widetilde{r}_{\alpha,I}(t)\,,
\end{align}
where $n_B(\omega) = \frac{1}{e^{\frac{\omega}{k_B {\rm T}}}-1}$ is the Bose-Einstein distribution of the bosonic bath at temperature T.  Hence, the average power becomes
\begin{align}
\langle\hat{P}\rangle&=\sum_\alpha\int_{0}^\infty d\tau\int_0^\infty d\omega\omega g_\alpha^2(\omega)\left[\chi_{\alpha,r}(t,t-\tau)[n_B(\omega)e^{-i\omega\tau}-(1+n_B(\omega))e^{i\omega\tau}]+(e^{-i\omega\tau}+e^{i\omega\tau})\chi_{\alpha,<}(t-\tau,t)\right]\,,
\end{align}
where we defined the retarded and the lesser susceptibilities, respectively:
\begin{align}
    \chi_{\alpha,r}(t,t')&=\theta(t-t')\langle[\widetilde{r}_{\alpha,I}(t),\widetilde{r}_{\alpha,I}(t')]\rangle\nonumber\,,\\
    \chi_{\alpha,<}(t,t')&=\langle \widetilde{r}_{\alpha,I}(t)\widetilde{r}_{\alpha,I}(t')\rangle\,.
\end{align}
If we are  interested on the average power over one period of time, we can define:
\begin{align}
    \langle\chi_{\alpha,<}(t-\tau,t)\rangle_T&=\frac{1}{T}\int_0^T dt\chi_{\alpha,<}(t-\tau,t)=\sum_{n,k} p_ne^{i(\epsilon_n-\epsilon_m+k\Omega)\tau}\widetilde{r}_\alpha^{nm}(k)\widetilde{r}_\alpha^{mn}(-k)\,,\\
        \langle\chi_{\alpha,r}(t,t-\tau)\rangle_T&=\frac{1}{T}\int_0^T dt\chi_{\alpha,r}(t,t-\tau)=\sum_{n,k} p_n[e^{-i(\epsilon_n-\epsilon_m-k\Omega)\tau}-e^{i(\epsilon_n-\epsilon_m+k\Omega)\tau}]\widetilde{r}_\alpha^{nm}(k)\widetilde{r}_\alpha^{mn}(-k)\nonumber\\
        &=-\sum_{n,k} (p_n-p_m)e^{i(\epsilon_n-\epsilon_m+k\Omega)\tau}\widetilde{r}_\alpha^{nm}(k)\widetilde{r}_\alpha^{mn}(-k)\,.
\end{align}
Therefore, the average power over a period $T$ becomes:
\begin{align}
\langle\hat{P}\rangle_T&=\sum_\alpha\int_0^\infty d\tau\int_0^\infty d\omega\omega g_\alpha^2(\omega)\sum_{n,m,k}\left[(p_n-p_m)[n_B(\omega)e^{-i\omega\tau}-(1+n_B(\omega))e^{i\omega\tau}]+p_n(e^{-i\omega\tau}+e^{i\omega\tau})\right]\nonumber\\
&\times e^{i(\epsilon_n-\epsilon_m+k\Omega)\tau}\widetilde{r}_\alpha^{nm}(k)\widetilde{r}_\alpha^{mn}(-k)\nonumber\\
&=\sum_{\alpha,n,m,k}(\epsilon_n-\epsilon_m+k\Omega)g_\alpha^2(\epsilon_n-\epsilon_m+k\Omega)\left[(p_n-p_m)(2n_B(\epsilon_n-\epsilon_m+k\Omega)+1)+p_n+p_m
\right]|\widetilde{r}_\alpha^{nm}(k)|^2\,.
\end{align}
It is important to mention that in the scenario where there is no external driving ($\Omega=0$) and the system is in equilibrium, $p_n=e^{-\beta \epsilon_n}/Z$, where $Z=\sum_ne^{-\beta \epsilon_n}$, results in $\langle\hat{P}\rangle=0$, which aligns with our expectations. 

To make further progress, let us assume that only the Floquet levels stemming from the originally (undriven) degenerate subspace are occupied. Consequently, we can write:
\begin{align}
\langle\hat{P}\rangle_T&\approx\left(\frac{\Omega \lambda_0}{\omega_0 \lambda_{SO}}\right)^2\Omega\sum_{\alpha,\tau,\tau',k}(\gamma_B^\tau-\gamma_B^{\tau'}+k)^3g_\alpha^2[(\gamma_B^\tau-\gamma_B^{\tau'}+k)\Omega]\nonumber\\
&\times\left[(p_\tau-p_{\tau'})\left(2n_B[(\gamma_B^\tau-\gamma_B^{\tau'}+k)\Omega]+1\right)+p_\tau+p_{\tau'}
\right]|m^\alpha_{\tau\tau'}(k)|^2\,.
\end{align}
To extract estimates, let us apply this expression to a specific environment and driving. Choosing an ohmic environment, with $g_\alpha^2(\omega)\equiv\rho(\omega)=\gamma_\Omega\omega$, as in the previous section, and linear driving, we find: 
\begin{align}
\langle\hat{P}\rangle_T&\approx2\gamma_\Omega\left(\frac{\Omega \lambda_0}{\omega_0 \lambda_{SO}}\right)^2\left(\frac{R_{0,y}}{\lambda_{SO}}\right)^2\Omega^2\equiv \Gamma_+\Omega\,.
\end{align}


\section{Comparison between the effective eigenstates and the full QD Floquet states}
\label{numerics}

\begin{figure}[h]
    \centering
    \includegraphics[scale=0.8]{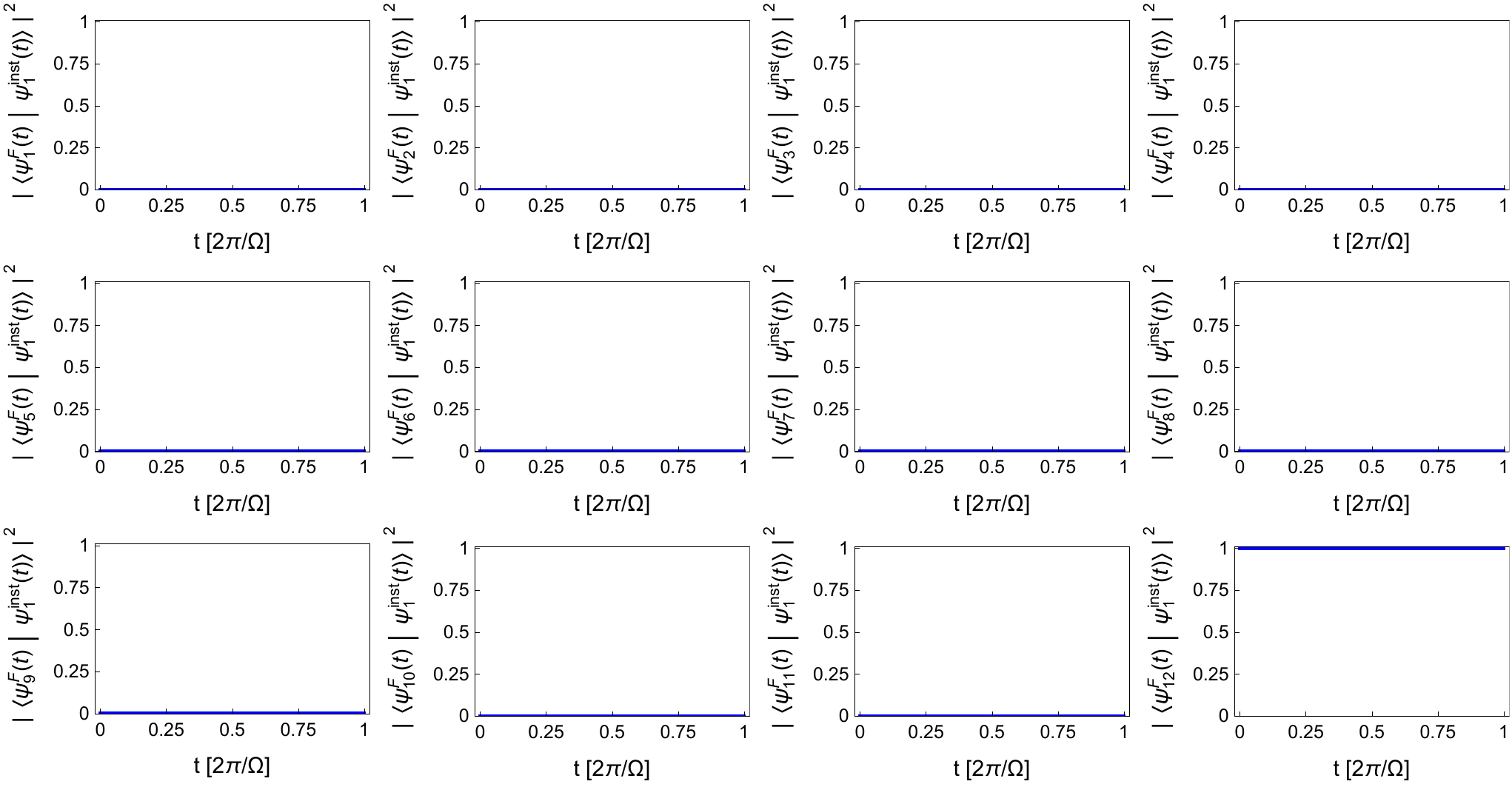}
    \caption{The overlap $|\langle \Psi_i^F(t)| \Psi^{\rm inst}_1(t) \rangle |^2$ of the fixed instantaneous state $| \Psi^{\rm inst}_1(t)\rangle$ with the Floquet states $\{|\Psi_i^F(t)\rangle \}_{i=1\dots i_{max}}$, evaluated numerically, with $i_{max}=12$. We observe that the Floquet state with index $i=12$, $| \Psi_{12}^F(t)\rangle $, exhibits nearly perfect overlap with $| \Psi^{\rm inst}_1(t) \rangle$, and hence the $| \Psi_{12}^F(t)\rangle $ state is selected as the first Floquet qubit state.  We assume a driven harmonic potential with parameters $E_0=0.1$, $\Omega=0.3$ and $\omega_0=1$, which pertain to the adiabatic regime defined in the main text.}
    \label{inst1}
\end{figure}
\begin{figure}
    \centering
    \includegraphics[scale=0.8]{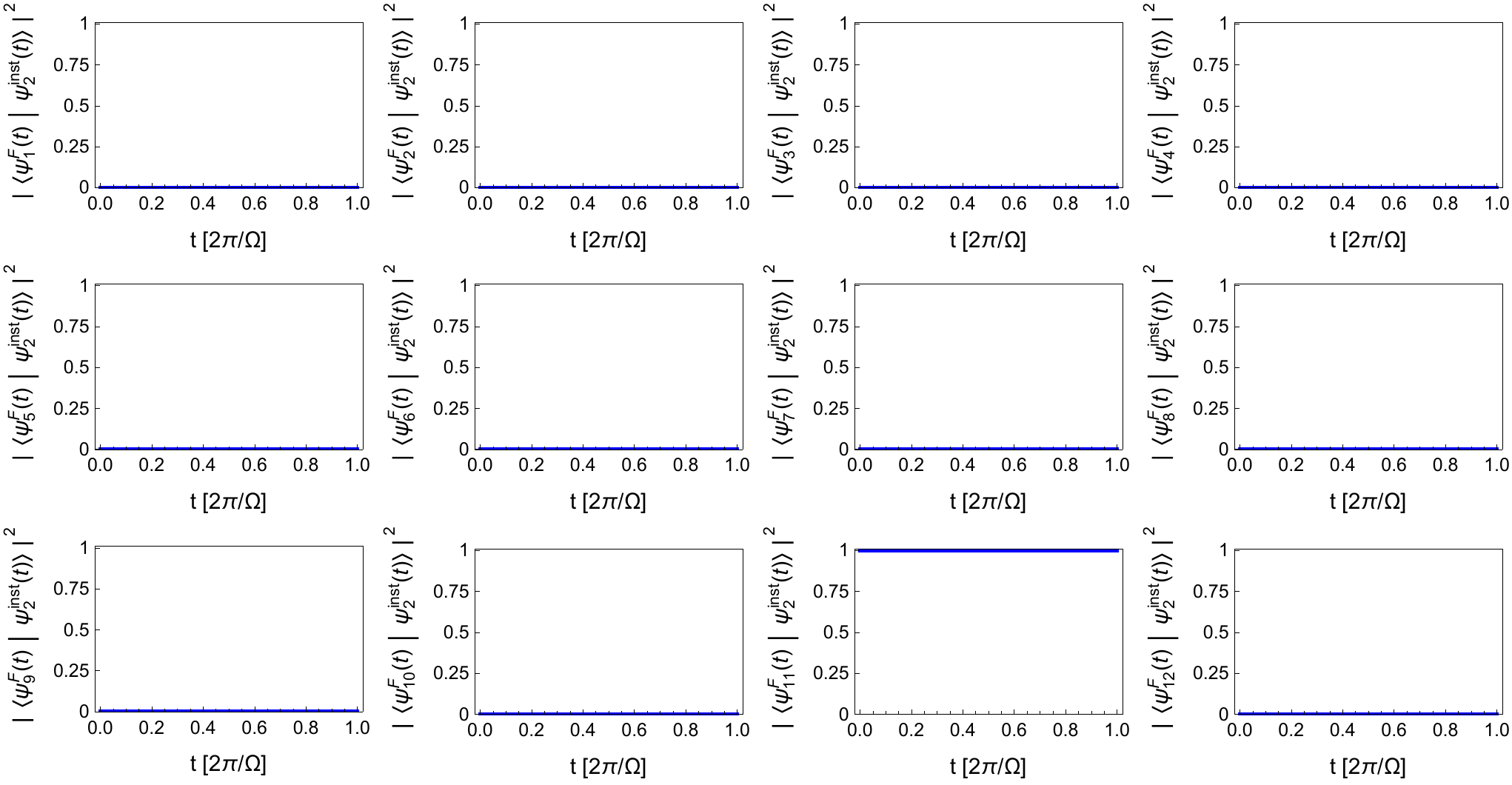}
    \caption{The overlap $|\langle \Psi_i^F(t)| \Psi^{\rm inst}_2(t) \rangle |^2$ of the fixed instantaneous state $| \Psi^{\rm inst}_2(t)\rangle$ with the Floquet states $\{|\Psi_i^F(t)\rangle \}_{i=1\dots i_{max}}$, evaluated numerically, with $i_{max}=12$. We observe that the Floquet state with index $i=12$, $| \Psi_{11}^F(t)\rangle $, exhibits a nearly perfect overlap with $| \Psi^{\rm inst}_1(t) \rangle$, and hence the $| \Psi_{11}^F(t)\rangle $ state is selected as the second Floquet qubit state. All other parameters are the same as in Fig.~\ref{inst1}.}
    \label{inst2}
\end{figure}
\end{widetext}

In this Appendix, we establish the validity of the low-energy model utilized in the manuscript to describe the effective Floquet dynamics. There, we assumed that the driving is adiabatic compared to the orbital level spacing and involves only the lowest degenerate states in the Floquet evolution of the QD. Here, we validate the accuracy of this approximation by conducting numerical simulations on the full QD Hamiltonian.  

We consider a QD confined by a harmonic confining potential of strength $\omega_0$. When the QD is subject to an electric field $\mathbf{E}(t)= E_0 \left( \cos(\Omega t), \sin(\Omega t)\right)$, it displaces the harmonic potential in a periodic manner. We have examined the Floquet spectrum and eigenvalues for different magnitudes of the electric field $E_0$ and frequencies $\Omega$.  However, in the context of the Floquet Hamiltonian we are considering, the quasienergies are only defined modulo $\text{Mod}[\Omega]$. As a result, we are unable to directly associate the lowest quasienergies with the two Floquet spin-qubit states. Instead, we must depend on the instantaneous eigenstates to identify them. Therefore, we evaluated  the inner product $|\langle\Psi_i^F(t)| \Psi^{\text{inst}}_j(t) \rangle |^2$ of a given instantaneous state in the lowest subspace $| \Psi^{\text{inst}}_j(t) \rangle $, with all the obtained Floquet states. Fig.~\eqref{inst1} shows that the Floquet state $|\Psi_{12}^F(t)\rangle $ overlaps with the instantaneous state $|\Psi^{\text{inst}}_{1}(t) \rangle$, while and Fig.~\ref{inst2} highlights that $|\Psi_{11}^F(t)\rangle $ overlaps with the other (i.e. Kramer pair) instantaneous state $|\Psi^{\text{inst}}_{2}(t) \rangle$. Now, by selecting these two Floquet states, $\{ \Psi_{11}^F(t)\rangle,\, \Psi_{12}^F(t)\rangle\}$, as the spin-qubit states, it becomes possible to carry out the qubit operations described in the text. For other potentials, the same procedure can be utilized to numerically determine the Floquet states and compare them with the lowest instantaneous states to uncover the Floquet spin-qubit states. 

\bibliography{ref} 
 
\end{document}